\documentclass[prd,showpacs,nofootinbib,preprintnumbers]{revtex4}
\usepackage{amsmath}
\usepackage{amsfonts}
\usepackage{graphicx}
\usepackage{dcolumn}
\usepackage{hyperref}

\def\be{\begin{equation}}
\def\ee{\end{equation}}
\def\bea{\begin{eqnarray}}
\def\eea{\end{eqnarray}}

\def\5{\overline 5}

\begin{document}

\title{Dark Energy Interacting with Dark Matter\\ in Classical Einstein and Loop Quantum Cosmology}

\author{Song Li and Yongge Ma}
\affiliation{Department of Physics, Beijing Normal University,
Beijing 100875, P. R. China}

\begin{abstract}
The cosmological model of dark energy interacting with cold dark
matter without coupling to the baryonic matter, is studied in the
background of both classical Einstein and loop quantum cosmology. We
consider two types of interacting models. In the former model, the
interaction is a linear combination of the densities of two dark
sectors, while in the latter model, the interaction with a constant
transfer rate depends only on the density of cold dark matter. It is
shown that the dynamical results in loop quantum cosmology are
different from those in classical Einstein cosmology for both two
kinds of interacting models. Moreover, the form of the interaction
affects significantly the dynamical results in both kinds of
cosmology.

\end{abstract}

\pacs{98.80.Cq, 98.80.-k}\maketitle

\section{Introduction}

Recently, the discovery of the acceleration of cosmological
expansion at present epoch has been the most principal achievement
of observational cosmology. Numerous cosmological observations, such
as Type Ia Supernovae (SNIa)~\cite{price1}, Cosmic Microwave
Background Radiation (CMBR)~\cite{price2} and Large Scale
Structure~\cite{price3}, strongly suggest that the universe is
spatially flat with about $4\%$ ordinary baryonic matter, $20\%$
dark matter and $76\%$ dark energy. The accelerated expansion of the
present universe is attributed to the dominant component of the
universe, dark energy, which has a large negative pressure but not
cluster. In fact, it has not been detected directly and there is no
justification for assuming that dark energy resembles known forms of
matter or energy. A large body of recent work has focussed on
understanding the nature of dark energy. However, the physical
origin of dark energy as well as its nature remain enigmatic at
present.

The simplest model of dark energy is the cosmological constant
$\Lambda$~\cite{price4}, whose energy density remains constant with
time $\rho_{\Lambda}=\Lambda/8\pi G$ (natural units $c=\hbar=1$ is
used throughout the paper) and whose equation of state (defined as
the ratio of pressure to energy density) remains $w=-1$ as the
universe evolves. Unfortunately, the model is burdened with the
well-known cosmological constant problems, namely the fine-tuning
problem: why is the energy of the vacuum so much smaller than its
estimation? and the cosmic coincidence problem: why is the dark
energy density approximately equal to the matter density today?
These problems have led many researchers to try different approaches
to the dark energy issue. A possible method is to assume the
equation of state (EoS) $w$ is a dynamical variable, and thus the
dynamical scenario of dark energy is investigated. The most popular
model among them is dubbed quintessence~\cite{price5}. Besides,
other scalar-field dark energy models have been studied, including
phantom~\cite{price6}, tachyon~\cite{price7}, quintom~\cite{price8},
ghost condensates~\cite{price9}, etc. Also, there are other
candidates, for example, Chaplygin gas which attempt to unify dark
energy and dark matter~\cite{price10}, braneworld
model~\cite{price11} and 5-dimensional gravity model~\cite{phys1}
which explain the acceleration through the assumption that spacetime
has five dimensions instead of the usual four. In addition, since
the cosmological scaling solution (i.e., the energy densities of
dark energy and cold dark matter remain proportional) could probably
alleviate the coincidence problem, interacting dark energy models
are also proposed~\cite{price12}.

As we all know, observations at the level of the solar system
severely constrain non-gravitational interactions of baryons,
namely, non-minimal coupling between dark energy and ordinary matter
fluids is strongly restricted by the experimental tests in the solar
systems~\cite{phys23}, we therefore neglect this possibility.
However, since the nature of dark sectors remains unknown, it is
possible to have non-gravitational interactions between dark energy
and dark matter. So we focus on dark energy interacting with dark
matter alone.

Actually, many dark energy models are considered in the framework of
classical Einstein cosmology. However, an outstanding problem in
classical Einstein cosmology is the big bang singularity which is
expected to be solved by quantum gravity. As a background
independent quantization of general relativity, loop quantum gravity
(LQG) is one of the best candidate theories of quantum
gravity~\cite{phys71}. It has been applied in cosmology to analyze
our universe, known as Loop Quantum Cosmology (LQC)~\cite{phys72}.
In LQC, non-perturbative effects lead to $-\rho^{2}/\rho_{c}$
corrections to the standard Friedmann equation and thus allow us the
possibility of resolving any past and future
singularities~\cite{phys72,phys73}. The modification becomes
important when energy density of the universe becomes to be the same
order of a critical density $\rho_{c}$. When the correction term
$-\rho^{2}/\rho_{c}$ dominates during the evolution of our universe,
it will cause the quantum bounce and hence avoid the singularity.
Recently, more and more researchers have taken their attention to
LQC for the appealing features: avoidance of various
singularities~\cite{phys74}, inflation in LQC~\cite{phys75}, large
scale effect~\cite{phys2} and so on. Concretely, some dark energy
models are investigated in the background of LQC, such as
phantom~\cite{phys76}, coupling phantom~\cite{phys77}, quintom and
hessence~\cite{phys78}, interacting dark energy model~\cite{phys79},
etc.

In this paper, we study the dynamical evolution of two classes of
interacting dark energy models in classical Einstein and Loop
Quantum Cosmology. Here some questions naturally arise as follows.
Can these models alleviate the coincidence problem in classical
Einstein cosmology? Are there scaling solutions arising from the
effect of loop quantum cosmology? Can the future singularities be
resolved in LQC? By our analysis, it turns out that in the former
model, there are two attractors in classical Einstein cosmology and
LQC. One is an accelerated scaling solution and the other is a
baryon dominant solution. However, in the latter model, there exists
one attractor in classical Einstein cosmology, which is a dark
energy dominated solution rather than a scaling solution, whereas in
LQC all fixed points are unstable. Thus, there exists no scaling
solution in the latter case, namely, this kind of interacting dark
energy model can not be regarded as a candidate to alleviate the
coincidence problem. Also, we find that dynamical results in LQC are
different from those in classical Einstein cosmology for both two
kinds of interacting models. Our universe finally enters an
oscillating phase in LQC. Moreover, the oscillating frequencies are
significantly different for varied parameters of models. These
results are different from the those obtained in classical Einstein
cosmology. Thus, LQC allow us the possibility of resolving future
singularities. Hence, the quantum gravity effect may be manifested
in large scale in the interacting dark energy models.

In Sec. II, we study dynamical properties for the general case in
classical Einstein cosmology and LQC. Then the dynamical results of
two types of interacting models are respectively studied in Secs.
III and IV. In Sec. V, the numerical results are presented. Finally,
the conclusions are summarized in Sec. VI.

\section{Interacting Dark Energy Model in Classical Einstein Cosmology and LQC}

For a spatially flat universe, the total energy conservation
equation is
\begin{equation}
\dot{\rho}+3H(\rho+p)=0,
\end{equation}
where $H$ is the Hubble parameter, $\rho$ is the total energy
density and $p$ is the total pressure of the background fluid.

In our scenario, the universe contains dark energy, cold dark matter
and baryonic matter. Moreover, the two dark sectors interact through
the interaction term $Q$ and the baryonic matter only interacts
gravitationally with the dark sectors. Then the energy conservation
equation is written as
\begin{equation}
\dot{\rho}_{b}+3H\rho_{b}=0,
\end{equation}
\begin{equation}
\dot{\rho}_{m}+3H\rho_{m}=Q,
\end{equation}
\begin{equation}
\dot{\rho}_{d}+3H(1+w_{d})\rho_{d}=-Q,
\end{equation}
where the subscripts $b$, $m$ and $d$ respectively denote baryonic
matter, cold dark matter and dark energy. In this paper, we consider
the simplest case of dark energy with constant equation of state
$w_{d}=p_{d}/\rho_{d}$~\cite{constant}, although the equation of
state for dark energy could also be dynamic. Thus, $Q$ denotes the
energy density exchange in the dark sectors and the sign of $Q$
determines the direction of energy transfer. A positive $Q$
corresponds to the transfer of energy from dark energy to dark
matter, while a negative $Q$ represents the other way round. Due to
the unknown nature of dark sectors, there is as yet no basis in
fundamental theory for a special coupling between two dark sectors.
So the interaction term $Q$ discussed currently have to be chosen in
a phenomenological way~\cite{phys40}. Since there is no clear
consensus on the form of the coupling, different versions, that
arise from a variety of motivations, coexist in the literature.

\subsection{Classical Einstein Cosmology}

In classical Einstein cosmology, the Friedmann equation is given by
\begin{equation}
H^{2}=\frac{\kappa^{2}}{3}\rho=\frac{\kappa^{2}}{3}(\rho_{d}+\rho_{m}+\rho_{b}),
\end{equation}
where $\kappa^{2}\equiv8\pi G$. Then differentiating the above
equation with respect to cosmic time $t$ and using the total energy
conservation equation, we can get the Raychaudhuri equation
\begin{eqnarray}
\dot{H}&=&-\frac{\kappa^{2}}{2}(\rho+p)\nonumber\\
&=&-\frac{\kappa^{2}}{2}((1+w_{d})\rho_{d}+\rho_{m}+\rho_{b}).
\end{eqnarray}

To analyze the evolution of the dynamical system, we introduce the
following set of dimensionless variables:
\begin{equation}
x\equiv\frac{\kappa^{2}\rho_{d}}{3H^{2}},~~~~~~~~y\equiv\frac{\kappa^{2}\rho_{m}}{3H^{2}},~~~~~~~~z\equiv\frac{\kappa^{2}\rho_{b}}{3H^{2}},~~~~~~~~u\equiv\frac{\kappa^{2}Q}{3H^{3}}.
\end{equation}
Accordingly, the Friedmann constraint is
\begin{equation}
x+y+z=1,
\end{equation}
and Eqs. (5) and (6) can be written as
\begin{equation}
-\frac{\dot{H}}{H^{2}}=\frac{3}{2}(1+w_{d}x).
\end{equation}
Furthermore, using these variables, the EoS of the total cosmic
fluid is given by
\begin{equation}
w=\frac{p}{\rho}=\frac{w_{d}x}{x+y+z}=w_{d}x.
\end{equation}
Then, inserting the expression (7) into Eqs.(2)-(6), we can obtain
the following autonomous system:
\begin{eqnarray}
x^{'}&=&-3w_{d}x(1-x)-u,\\
y^{'}&=&3w_{d}x y+u,\\
z^{'}&=&3w_{d}x z,
\end{eqnarray}
where the prime denotes a derivative with respect to $N\equiv \ln
a$. We set the current scale factor by $a_{0}=1$, then the current
value of $N$ reads $N_{0}=0$. Setting $x^{'}=y^{'}=z^{'}=0$, we can
find the general solution of the critical points
$(x_{*},y_{*},z_{*})$ of the autonomous system (11)-(13) as the type
of $(x_{*}=0,u_{*}=0)$ and $(x_{*}+y_{*}=1,z_{*}=0)$.

\subsection{Loop Quantum Cosmology}

Due to the quantum effects in LQC, we consider effective Friedmann
equation with correction of the form~\cite{phys72}
\begin{equation}
H^{2}=\frac{\kappa^{2}}{3}\rho(1-\frac{\rho}{\rho_{c}}),
\end{equation}
where $\rho=\rho_{d}+\rho_{m}+\rho_{b}$, the critical density
$\rho_{c}\equiv\frac{\sqrt{3}}{16\pi^{2}\gamma^{3}}\rho_{pl}$
measures the loop quantum effects, $\rho_{pl}$ is the Planck
density, $\gamma$ is the dimensionless Barbero-Immirzi
parameter~\cite{phys70}. An important feature for the modified
dynamics is that a $\rho^2$ term which is relevant in the high
energy regime is included in the classical Friedmann equation. The
correction term predicts a bounce when the matter energy density
reaches the critical value $\rho_{c}$ which is close to the Planck
density. By the numerical simulation~\cite{phys72}, it turns out
that the modified Friedmann equation is valid in the whole
evolutional trajectory of the universe including the bounce.
Additionally, along with the total energy conservation equation, we
get
\begin{eqnarray}
\dot{H}&=&-\frac{\kappa^{2}}{2}(\rho+p)(1-2\frac{\rho}{\rho_{c}})\nonumber\\
&=&-\frac{\kappa^{2}}{2}((1+w_{d})\rho_{d}+\rho_{m}+\rho_{b})(1-2\frac{\rho_{d}+\rho_{m}+\rho_{b}}{\rho_{c}}).
\end{eqnarray}

Using the dimensionless variables defined in (7), the Friedmann
constraint is
\begin{equation}
(x+y+z)(1-\frac{3H^{2}}{\kappa^{2}}\frac{x+y+z}{\rho_{c}})=1,
\end{equation}
and Eqs. (14) and (15) can be written as
\begin{equation}
-\frac{\dot{H}}{H^{2}}=\frac{3}{2}(2-x-y-z)(1+\frac{w_{d}x}{x+y+z}).
\end{equation}
Furthermore, the EoS of the total cosmic fluid reads
\begin{equation}
w=\frac{p}{\rho}=\frac{w_{d}x}{x+y+z}.
\end{equation}
Then, Eqs.(2)-(4) combined with Eqs. (14)-(15) can be rewritten as
the following autonomous system according to the expression (7),
\begin{eqnarray}
x^{'}&=&-3(1+w_{d})x-u+3x(2-x-y-z)(1+\frac{w_{d}x}{x+y+z}),\\
y^{'}&=&-3y+u+3y(2-x-y-z)(1+\frac{w_{d}x}{x+y+z}),\\
z^{'}&=&-3z[1-(2-x-y-z)(1+\frac{w_{d}x}{x+y+z})].
\end{eqnarray}
The type of critical points of the autonomous system (19)-(21) can
be summarized as $(x_{*}=0,u_{*}=0)$, $(x_{*}+y_{*}=1,z_{*}=0)$ and
$(x_{*}\neq0,y_{*}=-(1+w_{d})x_{*},u=-3(1+w_{d})x_{*})$.

\section{Interacting Dark Energy Model I}

Let us first consider the interaction term
$Q=3H(c_{1}\rho_{m}+c_{2}\rho_{d})$~\cite{phys41,phys60}, where
$c_{1}$ and $c_{2}$ are coupling constants. Note that this form,
which was first proposed in~\cite{phys42}, is more general than
those proposed in~\cite{phys40,phys43}. The latter can be obtained
from the former by setting $c_{1}=c_{2}=c$, $c_{1}=0$ or $c_{2}=0$.
According to Ref.~\cite{phys41}, we assume the coupling constants
$c_{1}$ and $c_{2}$ have the same sign to achieve a physically
viable model.

\begin{table}[tbp]
\begin{center}
\begin{tabular}{|c||c|c|c|}
  \toprule
  Point & $(x_{*},y_{*},z_{*})$ & Eigenvalues & $w_{*}$
\\\hline\hline
  $~$ & ~ & $\lambda_{1}=3x_{s}$, & ~
  \\
  $A$ & $(-\frac{c_{1}-c_{2}-w_{d}-x_{s}}{2w_{d}},-\frac{c_{1}-c_{2}+w_{d}-x_{s}}{2w_{d}},0)$ & $\lambda_{2}=\lambda_{3}=-\frac{3}{2}(c_{1}-c_{2}-w_{d}-x_{s})$ & $-\frac{c_{1}-c_{2}-w_{d}-x_{s}}{2}$
  \\
  $~$ & ~ & ~ & ~
  \\\hline
  $~$ & ~ & $\lambda_{1}=-3x_{s}$, & ~
  \\
  $B$ & $(-\frac{c_{1}-c_{2}-w_{d}+x_{s}}{2w_{d}},-\frac{c_{1}-c_{2}+w_{d}+x_{s}}{2w_{d}},0)$ & $\lambda_{2}=\lambda_{3}=-\frac{3}{2}(c_{1}-c_{2}-w_{d}+x_{s})$ & $-\frac{c_{1}-c_{2}-w_{d}+x_{s}}{2}$
  \\
  $~$ & ~ & ~ & ~
  \\\hline
  $~$ & ~ & $\lambda_{1}=0$, & ~
  \\
  $C$ & $(0,0,1)$ & $\lambda_{2,3}=\frac{3}{2}(c_{1}-c_{2}-w_{d}\pm x_{s})$ &
  0
  \\
  \botrule
\end{tabular}
\end{center}
\caption{The properties of the critical points for the interacting
model I in classical Einstein cosmology. Here, the parameter $x_{s}$
is defined in Eq. (25).}
\end{table}

\subsection{Cosmological Dynamics in Classical Einstein Cosmology}

In classical Einstein cosmology, the autonomous system (11)-(13) can
be written as
\begin{eqnarray}
x^{'}&=&-3[(1+c_{2}+w_{d})x+c_{1}y]+3x(1+w_{d}x),\\
y^{'}&=&-3[(1-c_{1})y-c_{2}x]+3y(1+w_{d}x),\\
z^{'}&=&3w_{d}x z.
\end{eqnarray}
Furthermore, we can obtain the critical points $(x_{*},y_{*},z_{*})$
of the autonomous system as follows:
\begin{itemize}
  \item \textbf{Point A}:~~~
($-\frac{c_{1}-c_{2}-w_{d}-x_{s}}{2w_{d}}$,~~$-\frac{c_{1}-c_{2}+w_{d}-x_{s}}{2w_{d}}$,~~$0$),\\
  \item \textbf{Point B}:~~~
($-\frac{c_{1}-c_{2}-w_{d}+x_{s}}{2w_{d}}$,~~$-\frac{c_{1}-c_{2}+w_{d}+x_{s}}{2w_{d}}$,~~$0$),\\
  \item \textbf{Point C}:~~~
($0$,~~$0$,~~$1$).
\end{itemize}
Here the parameter $x_{s}$ is defined by
\begin{equation}
x_{s}=\sqrt{(c_{1}-c_{2}-w_{d})^{2}+4w_{d}c_{1}}.
\end{equation}

To study the stability of the critical points for the autonomous
system, we substitute linear perturbations $x\rightarrow
x_{*}+\delta x$, $y\rightarrow y_{*}+\delta y$ and $z\rightarrow
z_{*}+\delta z$ about the critical points into the autonomous system
Eqs. (22)-(24). To first-order in the perturbations, we get the
following evolution equations of the linear perturbations:
\begin{eqnarray}
\delta x^{'}&=&-3(c_{2}+w_{d}-2w_{d}x_{*})\delta x-3c_{1}\delta y,\\
\delta y^{'}&=&3(c_{2}+w_{d}y_{*})\delta x+3(c_{1}+w_{d}x_{*})\delta y,\\
\delta z^{'}&=&3w_{d}z_{*}\delta x+3w_{d}x_{*}\delta z.
\end{eqnarray}
The three eigenvalues of the coefficient matrix of Eqs. (26)-(28)
determine the stability of the critical points. We list the three
eigenvalues for each point in Table I. We examine the sign of the
eigenvalues of points A and B and find that point A is not stable if
it exists, whereas point B is stable if $0<c_{1}<-w_{d}$ and
$0<c_{2}<c_{1}-w_{d}-2\sqrt{-w_{d}c_{1}}$, i.e., the critical point
B is always the stable attractor solution if it exists. In addition,
at point B, from the expression of the total EoS
$w_{*}=-\frac{c_{1}-c_{2}-w_{d}+x_{s}}{2}$, the acceleration
condition $w<-1/3$ shows that if
\begin{eqnarray}
\left\{\begin{array}{lll}
w_{d}>-1/3, &~~0<c_{1}<-w_{d},& ~~0<c_{2}<(1+3w_{d})c_{1}-w_{d}+1/3;\\
-2/3<w_{d}<-1/3,&~~ w_{d}+2/3<c_{1}<-1/(9w_{d}),&~~0<c_{2}<(1+3w_{d})c_{1}-w_{d}+1/3;\\
w_{d}<-2/3,&~~0<c_{1}<-1/(9w_{d}),& ~~0<c_{2}<(1+3w_{d})c_{1}-w_{d}+1/3;\\
w_{d}<-1/3,&~~-1/(9w_{d})<c_{1}<-w_{d},&~~0<c_{2}<c_{1}-w_{d}-2\sqrt{-w_{d}c_{1}},
\end{array}\right.
\end{eqnarray}
point B is an accelerated scaling attractor where the energy
densities of dark energy and cold dark matter remain proportional.
Thus it may alleviate the coincidence problem. The baryon dominated
point C is stable if $c_{1}-c_{2}-w_{d}<0$ and $4w_{d}c_{1}<0$. This
condition corresponds to $c_{1}>0$ and $c_{2}>c_{1}-w_{d}$ under the
prior condition $w_{d}<0$.

\begin{table}[tbp]
\begin{center}
\begin{tabular}{|c||c|c|c|}
  \toprule
  Point & $(x_{*},y_{*},z_{*})$ & Eigenvalues & $w_{*}$
\\\hline\hline
  $~$ & ~ & $\lambda_{1}=3x_{s}$, & ~
  \\
  $A$ & $(-\frac{c_{1}-c_{2}-w_{d}-x_{s}}{2w_{d}},-\frac{c_{1}-c_{2}+w_{d}-x_{s}}{2w_{d}},0)$ & $\lambda_{2}=-\frac{3}{2}(c_{1}-c_{2}-w_{d}-x_{s})$, & $-\frac{c_{1}-c_{2}-w_{d}-x_{s}}{2}$
  \\
  $~$ & ~ & $\lambda_{3}=-\frac{3}{2}(2-c_{1}+c_{2}+w_{d}+x_{s})$ & ~
  \\\hline
  $~$ & ~ & $\lambda_{1}=-3x_{s}$, & ~
  \\
  $B$ & $(-\frac{c_{1}-c_{2}-w_{d}+x_{s}}{2w_{d}},-\frac{c_{1}-c_{2}+w_{d}+x_{s}}{2w_{d}},0)$ & $\lambda_{2}=-\frac{3}{2}(c_{1}-c_{2}-w_{d}+x_{s})$, & $-\frac{c_{1}-c_{2}-w_{d}+x_{s}}{2}$
  \\
  $~$ & ~ & $\lambda_{3}=-\frac{3}{2}(2-c_{1}+c_{2}+w_{d}-x_{s})$ & ~
  \\\hline
  $~$ & ~ & $\lambda_{1}=-3$, & ~
  \\
  $C$ & $(0,0,1)$ & $\lambda_{2,3}=\frac{3}{2}(c_{1}-c_{2}-w_{d}\pm x_{s})$ &
  0
  \\
  \botrule
\end{tabular}
\end{center}
\caption{The properties of the critical points for the interacting
model I in LQC. Also, the parameter $x_{s}$ is defined in Eq. (25).}
\end{table}

\subsection{Cosmological Dynamics in LQC}

We consider the autonomous system (19)-(21) in LQC. By inserting the
concrete form of $Q$ into Eqs.(19)-(21), the autonomous system can
be expressed as
\begin{eqnarray}
x^{'}&=&-3[(1+c_{2}+w_{d})x+c_{1}y]+3x(2-x-y-z)(1+\frac{w_{d}x}{x+y+z}),\\
y^{'}&=&-3[(1-c_{1})y-c_{2}x]+3y(2-x-y-z)(1+\frac{w_{d}x}{x+y+z}),\\
z^{'}&=&-3z[1-(2-x-y-z)(1+\frac{w_{d}x}{x+y+z})].
\end{eqnarray}
The corresponding critical points $(x_{*},y_{*},z_{*})$ of the
autonomous system (30)-(32) are obtained as follows:
\begin{itemize}
  \item \textbf{Point A}:~~~
($-\frac{c_{1}-c_{2}-w_{d}-x_{s}}{2w_{d}}$,~~$-\frac{c_{1}-c_{2}+w_{d}-x_{s}}{2w_{d}}$,~~$0$),\\
  \item \textbf{Point B}:~~~
($-\frac{c_{1}-c_{2}-w_{d}+x_{s}}{2w_{d}}$,~~$-\frac{c_{1}-c_{2}+w_{d}+x_{s}}{2w_{d}}$,~~$0$),\\
  \item \textbf{Point C}:~~~
($0$,~~$0$,~~$1$).
\end{itemize}

In order to study the stability of the critical points for the
autonomous system (30)-(32), we obtain the following evolution
equations of the linear perturbations:
\begin{eqnarray}
\delta x^{'}&=&-3[-1+c_{2}+w_{d}+2x_{*}+y_{*}+z_{*}+\frac{w_{d}x_{*}^{2}}{x_{*}+y_{*}+z_{*}}-\frac{w_{d}x_{*}(2-x_{*}-y_{*}-z_{*})(x_{*}+2y_{*}+2z_{*})}{(x_{*}+y_{*}+z_{*})^{2}}]\delta x\nonumber\\
&&-3[c_{1}+x_{*}+\frac{2w_{d}x_{*}^{2}}{(x_{*}+y_{*}+z_{*})^{2}}]\delta y-3x_{*}[1+\frac{2w_{d}x_{*}}{(x_{*}+y_{*}+z_{*})^{2}}]\delta z,\\
\delta y^{'}&=&3[c_{2}-(1+w_{d})y_{*}+\frac{2w_{d}y_{*}(y_{*}+z_{*})}{(x_{*}+y_{*}+z_{*})^{2}}]\delta x\nonumber\\
&&+3[1+c_{1}-(1+w_{d})x_{*}-2y_{*}-z_{*}+\frac{2w_{d}x_{*}(x_{*}+z_{*})}{(x_{*}+y_{*}+z_{*})^{2}}]\delta y-3y_{*}[1+\frac{2w_{d}x_{*}}{(x_{*}+y_{*}+z_{*})^{2}}]\delta z,\\
\delta
z^{'}&=&-3z_{*}[1+w_{d}-\frac{2w_{d}(y_{*}+z_{*})}{(x_{*}+y_{*}+z_{*})^{2}}]\delta
x\nonumber\\
&&-3z_{*}[1+\frac{2w_{d}x_{*}}{(x_{*}+y_{*}+z_{*})^{2}}]\delta
y-3[-1+(1+w_{d})x_{*}+y_{*}+2z_{*}-\frac{2w_{d}x_{*}(x_{*}+y_{*})}{(x_{*}+y_{*}+z_{*})^{2}}]\delta
z.
\end{eqnarray}
The corresponding three eigenvalues of the coefficient matrix of
Eqs.(33)-(35) are listed in Table II. It is easy to examine that
point A is not a stable point, while point B is stable if $w_{d}>-1,
0<c_{1}<-w_{d}, 0<c_{2}<c_{1}-w_{d}-2\sqrt{-w_{d}c_{1}}$ or
$w_{d}<-1, 0<c_{1}<-1/w_{d},
(1+w_{d})(c_{1}-1)<c_{2}<c_{1}-w_{d}-2\sqrt{-w_{d}c_{1}}$. We find
that for the case of $w>-1$, the stable regions of parameters in LQC
are the same as those in classical Einstein cosmology. Additionally,
at point B, the accelerated condition $w<-1/3$ can be expressed as

\begin{eqnarray}
\left\{\begin{array}{lll}
0<c_{1}<-w_{d},&~~0<c_{2}<(1+w_{d})c_{1}-w_{d}+1/3 &~~\textrm{if $w_{d}>-1/3$};\\
w_{d}+2/3<c_{1}<-1/(9w_{d}),&~~ 0<c_{2}<(1+w_{d})c_{1}-w_{d}+1/3 &~~\textrm{if $-2/3<w_{d}<-1/3$};\\
0<c_{1}<-1/(9w_{d}),&~~ 0<c_{2}<(1+3w_{d})c_{1}-w_{d}+1/3 &~~\textrm{if $-1<w_{d}<-2/3$};\\
-1/(9w_{d})<c_{1}<-w_{d},&~~0<c_{2}<c_{1}-w_{d}-2\sqrt{-w_{d}c_{1}} &~~\textrm{if $-1<w_{d}<-1/3$};\\
-1/(9w_{d})<c_{1}<-1/(3w_{d}),&~~ (1+w_{d})(c_{1}-1)<c_{2}<c_{1}-w_{d}-2/3 &~~\textrm{if $w_{d}<-1$};\\
0<c_{1}<-1/(9w_{d}),&~~(1+w_{d})(c_{1}-1)<c_{2}<(1+3w_{d})c_{1}-w_{d}+1/3
&~~\textrm{if $w_{d}<-1$}
\end{array}\right.
\end{eqnarray}
so that point B is an accelerated scaling attractor for any case
above, which provides a possibility to alleviate the coincidence
problem. It should be noted that point C is stable for the same
condition as the case in the classical Einstein cosmology, and
therefore, we will neglect discussing this point in the section V.

\section{Interacting Dark Energy Model II}

In this section, we study the interaction dark energy model with a
constant transfer rate: $Q=\Gamma\rho_{m}$~\cite{phys50}, which has
been used in reheating~\cite{phys51}, dark matter
decay~\cite{phys52}, curvaton decay~\cite{phys53} and the decay of
superheavy dark matter particles to a quintessence scalar
field~\cite{phys54}.

In order to study the dynamical evolution, we additionally define a
new dimensionless variable as
\begin{equation}
v\equiv\frac{H_{0}}{H},
\end{equation}
where $H_{0}$ denotes the current value of the Hubble parameter, and
for convenience, we introduce the parameter
\begin{equation}
\beta=\frac{\Gamma}{H_{0}}.
\end{equation}

\begin{table}[tbp]
\begin{center}
\begin{tabular}{|c||c|c|c|}
  \toprule
  Point & $(x_{*},y_{*},z_{*},v_{*})$ & Eigenvalues & $w_{*}$
\\\hline\hline
  $~$ & ~ & $\lambda_{1}=\lambda_{2}=0$, & ~
  \\
  $A$ & $(0,y_{*},1-y_{*},0)$ & $\lambda_{3}=\frac{3}{2}$, $\lambda_{4}=-3w_{d}$ & $0$
  \\
  $~$ & ~ & ~ & ~
  \\\hline
  $~$ & ~ & $\lambda_{1}=\lambda_{2}=\lambda_{3}=3w_{d}$, & ~
  \\
  $B$ & $(1,0,0,0)$ & $\lambda_{4}=\frac{3}{2}(1+w_{d})$, & $w_{d}$
  \\
  $~$ & ~ & $~$ & ~
  \\\hline
  $~$ & ~ & $\lambda_{1}=\lambda_{2}=-3$, & ~
  \\
  $C$ & $(-\frac{1}{w_{d}},\frac{1+w_{d}}{w_{d}},0,\frac{3}{\beta})$ & $\lambda_{3,4}=\frac{3}{2}(-1-w_{d}\pm\sqrt{w_{d}^{2}-1})$ & $-1$
  \\
  $~$ & ~ & $~$ & ~
  \\
  \botrule
\end{tabular}
\end{center}
\caption{The properties of the critical points for the interacting
model II in classical Einstein cosmology.}
\end{table}

\subsection{Cosmological Dynamics in Classical Einstein Cosmology}

In classical Einstein cosmology, using Eqs.(2)-(6), (37) and (38),
we concretely express the autonomous system (11)-(13) as
\begin{eqnarray}
x^{'}&=&-3w_{d}x(1-x)-\beta v y,\\
y^{'}&=&-3w_{d}x y+\beta v y,\\
z^{'}&=&3w_{d}x z,\\
v^{'}&=&\frac{3}{2}v(1+w_{d}x),
\end{eqnarray}
which has three critical points as follows:
\begin{itemize}
  \item \textbf{Point A}:~~~
($0$,~~$y_{*}$,~~$1-y_{*}$,~~$0$),\\
  \item \textbf{Point B}:~~~
($1$,~~$0$,~~$0$,~~$0$),\\
  \item \textbf{Point C}:~~~
($-\frac{1}{w_{d}}$,~~$\frac{1+w_{d}}{w_{d}}$,~~$0$,~~$\frac{3}{\beta}$).
\end{itemize}

Substituting linear perturbations $x\rightarrow x_{*}+\delta x$,
$y\rightarrow y_{*}+\delta y$, $z\rightarrow z_{*}+\delta z$ and
$v\rightarrow v_{*}+\delta v$ about the critical points into the
autonomous system Eqs.(39)-(42), to first-order in the
perturbations, we get the following evolution equations of the
linear perturbations:
\begin{eqnarray}
\delta x^{'}&=&-3w_{d}(1-2x_{*})\delta x-\beta v_{*}\delta y-\beta y_{*}\delta v,\\
\delta y^{'}&=&3w_{d}y_{*}\delta x+(3w_{d}x_{*}+\beta v_{*})\delta y+\beta y_{*}\delta v,\\
\delta z^{'}&=&3w_{d}z_{*}\delta x+3w_{d}x_{*}\delta z,\\
\delta v^{'}&=&\frac{3}{2}w_{d}v_{*}\delta
x+\frac{3}{2}(1+w_{d}x_{*})\delta v.
\end{eqnarray}
The four eigenvalues of the coefficient matrix of the above
equations determine the stability of the critical points. In Table
III, we list the eigenvalues for each point. Then, it is clear that
the critical points A and C are not stable if they exist, while
point B is stable when $w_{d}<-1$. Since the total EoS at point B is
$w_{*}=w_{d}$, point B is an accelerated attractor. Note that it is
a dark energy dominated solution, rather than a scaling solution.
Thus, this kind of interacting model in classical Einstein cosmology
can not be regarded as a candidate to alleviate the coincidence
problem.

\begin{table}[tbp]
\begin{center}
\begin{tabular}{|c||c|c|c|}
  \toprule
  Point & $(x_{*},y_{*},z_{*},v_{*})$ & Eigenvalues & $w_{*}$
\\\hline\hline
  $~$ & ~ & $\lambda_{1}=0$, $\lambda_{2}=-3$ & ~
  \\
  $A$ & $(0,y_{*},1-y_{*},0)$ & $\lambda_{3}=\frac{3}{2}$, $\lambda_{4}=-3w_{d}$ & $0$
  \\
  $~$ & ~ & ~ & ~
  \\\hline
  $~$ & ~ & $\lambda_{1}=\lambda_{2}=3w_{d}$, & ~
  \\
  $B$ & $(1,0,0,0)$ & $\lambda_{3}=\frac{3}{2}(1+w_{d})$, $\lambda_{4}=-3(1+w_{d})$, & $w_{d}$
  \\
  $~$ & ~ & $~$ & ~
  \\\hline
  $~$ & ~ & $\lambda_{1}=0$, $\lambda_{2}=\frac{3(w_{d}+2)}{w_{d}}$ & ~
  \\
  $C$ & $(x_{*},-(1+w_{d})x_{*},0,\frac{3}{\beta})$ & $\lambda_{3,4}=\frac{3}{2}\{-1-w_{d}\pm\sqrt{(1+w_{d})[-3+w_{d}(1-2x_{*})]}\}$ & $-1$
  \\
  $~$ & ~ & $~$ & ~
  \\
  \botrule
\end{tabular}
\end{center}
\caption{The properties of the critical points for the interacting
model II in LQC.}
\end{table}

\subsection{Cosmological Dynamics in LQC}

In LQC, inserting Eqs.(37)and (38) into the evolution equations
(19)-(21) and making use of Eqs.(14) and (15), we get the autonomous
system
\begin{eqnarray}
x^{'}&=&-3(1+w_{d})x-\beta v y+3x(2-x-y-z)(1+\frac{w_{d}x}{x+y+z}),\\
y^{'}&=&-3y+\beta v y+3y(2-x-y-z)(1+\frac{w_{d}x}{x+y+z}),\\
z^{'}&=&-3z[1-(2-x-y-z)(1+\frac{w_{d}x}{x+y+z})],\\
v^{'}&=&\frac{3}{2}v(2-x-y-z)(1+\frac{w_{d}x}{x+y+z}).
\end{eqnarray}
The critical points of the autonomous system (47)-(50) is obtained
as
\begin{itemize}
  \item \textbf{Point A}:~~~
($0$,~~$y_{*}$,~~$1-y_{*}$,~~$0$),\\
  \item \textbf{Point B}:~~~
($1$,~~$0$,~~$0$,~~$0$),\\
  \item \textbf{Point C}:~~~
($x_{*}$,~~$-(1+w_{d})x_{*}$,~~$0$,~~$\frac{3}{\beta}$).
\end{itemize}
In order to study the stability of the critical points for the
autonomous system (47)-(50), we obtain the following evolution
equations of the linear perturbations:
\begin{eqnarray}
\delta x^{'}&=&-3[-1+w_{d}+2x_{*}+y_{*}+z_{*}+\frac{w_{d}x_{*}^{2}}{x_{*}+y_{*}+z_{*}}-\frac{w_{d}x_{*}(2-x_{*}-y_{*}-z_{*})(x_{*}+2y_{*}+2z_{*})}{(x_{*}+y_{*}+z_{*})^{2}}]\delta x\nonumber\\
&&-\{\beta v_{*}+3x_{*}[1+\frac{2w_{d}x_{*}}{(x_{*}+y_{*}+z_{*})^{2}}]\}\delta y-3x_{*}[1+\frac{2w_{d}x_{*}}{(x_{*}+y_{*}+z_{*})^{2}}]\delta z-\beta y_{*}\delta v,\\
\delta
y^{'}&=&-3y_{*}[1+w_{d}-\frac{2w_{d}(y_{*}+z_{*})}{(x_{*}+y_{*}+z_{*})^{2}}]\delta
x\nonumber\\
&&+\{\beta
v_{*}+3[1-(1+w_{d})x_{*}-2y_{*}-z_{*}+\frac{2w_{d}x_{*}(x_{*}+z_{*})}{(x_{*}+y_{*}+z_{*})^{2}}]\}\delta
y-3y_{*}[1+\frac{2w_{d}x_{*}}{(x_{*}+y_{*}+z_{*})^{2}}]\delta z+\beta y_{*}\delta v,\\
\delta
z^{'}&=&-3z_{*}[1+w_{d}-\frac{2w_{d}(y_{*}+z_{*})}{(x_{*}+y_{*}+z_{*})^{2}}]\delta x\nonumber\\
&&-3z_{*}[1+\frac{2w_{d}x_{*}}{(x_{*}+y_{*}+z_{*})^{2}}]\delta y-3[-1+(1+w_{d})x_{*}+y_{*}+2z_{*}-\frac{2w_{d}x_{*}(x_{*}+y_{*})}{(x_{*}+y_{*}+z_{*})^{2}}]\delta z,\\
\delta
v^{'}&=&-\frac{3}{2}v_{*}[1+w_{d}-\frac{2w_{d}(y_{*}+z_{*})}{(x_{*}+y_{*}+z_{*})^{2}}]\delta
x\nonumber\\
&&-\frac{3}{2}v_{*}[1+\frac{2w_{d}x_{*}}{(x_{*}+y_{*}+z_{*})^{2}}]\delta
y-\frac{3}{2}v_{*}[1+\frac{2w_{d}x_{*}}{(x_{*}+y_{*}+z_{*})^{2}}]\delta
z+\frac{3}{2}(2-x-y-z)(1+\frac{w_{d}x}{x+y+z})\delta v.
\end{eqnarray}
Solving the four eigenvalues of the coefficient matrix of the above
equations, we list them in Table IV. It is not difficult to see that
the critical point A is not stable. For point C, to guarantee the
energy densities of dark sectors to be positive, we get $w_{d}<-1$,
and therefore the real parts of the eigenvalues $\lambda_{3}$ and
$\lambda_{4}$ are positive. This means that point C is not stable.
The critical point B is also not stable, since the sign of
$\lambda_{3}$ is always opposite to the sign of $\lambda_{4}$.
However, in classical Einstein cosmology point B is stable when
$w_{d}<-1$.

\section{Numerical Results}

In what follows, we numerically study the dynamical results of the
interacting dark energy models to confirm the complicated stability
condition for the critical points in both interacting models.

\subsection{The Interacting Model I}

In the former model, there are two attractors in both classical
Einstein cosmology and LQC. One is a baryon dominated solution which
is neglected, and the other is an accelerated scaling solution.
Numerical results for the interacting model I are shown in Figs.1-8.

In Fig.1, we depict the parameter space $(c_{1},c_{2})$ to be stable
by choosing $w_{d}=-0.6$ and $w_{d}=-1.4$. When $w_{d}>-1$, the
stable region in LQC is the same as that in classical Einstein
cosmology, i.e., the critical point B is stable in the region I in
both kinds of cosmology. However, when $w_{d}<-1$, point B is stable
in the region I+II in classical Einstein cosmology, whereas in LQC,
point B is an attractor only in the region II.

In Fig.2, we plot the phase space trajectories of the universe with
$w_{d}$, $c_{1}$ and $c_{2}$ in the stable region. We find that the
position of the critical point B depends on the EoS $w_{d}$ and the
coupling constants $c_{1}$ and $c_{2}$, but is independent of the
theory describing our universe. However, the trajectories in the
phase space depend not only upon $w_{d}$, $c_{1}$ and $c_{2}$, but
also upon the selected theory.

Fig.3 shows the evolution of the total EoS $w$ with the chosen
parameters $w_{d}$, $c_{1}$ and $c_{2}$, satisfying the acceleration
conditions (29) and (36). Apparently, we can see that in the final
state the total EoS $w$ tends to a constant, which depends on
$w_{d}$, $c_{1}$ and $c_{2}$, but is independent of the theory
describing our universe.

Fig.4 exhibits the trajectories of scalar factor $a$ versus time for
different values of parameters in LQC. We set $\kappa^{2}=1$, and
thus take $\rho_{c}=1$ since the value of $\rho_{c}$ is on the order
of the Plank density, $\kappa^{-4}$. From the figure, one can see
that the evolution trajectories are significantly different for
varied parameters. The bounce in scale factor occurs later for
greater value of $w_{d}$, $c_{1}$ or $c_{2}$. Our universe finally
enters an oscillating phase in LQC.

In Figs.5-8, we plot the evolution trajectories of the Hubble
parameter $H$ and energy density $\rho$ versus time. The parameters
we selected is in the unstable region in LQC. Differentiating
Eq.(14) with respect to $\rho$, we find that $H$ has a extremum
value ($dH/d\rho=0$) when $\rho=\rho_{c}/2$. Additionally, the
second order derivative of $H$ reads
\begin{eqnarray}
&&(\frac{d^{2}H}{d\rho^{2}})_{\rho=\rho_{c}/2}=-\frac{2\kappa}{\sqrt{3\rho_{c}^{3}}}<0~~~~~~(H>0),\\
&&(\frac{d^{2}H}{d\rho^{2}})_{\rho=\rho_{c}/2}=\frac{2\kappa}{\sqrt{3\rho_{c}^{3}}}>0~~~~~~(H<0).
\end{eqnarray}
Thus, $H_{max}=\sqrt{\kappa^{2}\rho_{c}/12}$ when $H>0$, while
$H_{min}=-\sqrt{\kappa^{2}\rho_{c}/12}$ when $H<0$. In Figs.5-8,
with $\rho_{c}=1$, a calculation gives $H_{max,min}\approx\pm0.2887$
at $\rho=\rho_{c}/2$. When $H\approx0$, we have the density
$\rho\approx\rho_{c}$ and thus the bounce occurs. From the figures,
we find that the expansion of our universe halts at the time when
$H\approx0$, and then contracts until $H\approx0$ again. The
universe goes on bouncing forward and backward. It is worthwhile to
note that the oscillating frequencies of $H(t)$ and $\rho(t)$ depend
upon the chosen coupling constants $c_{1}$, $c_{2}$ and EoS $w_{d}$.

\subsection{The Interacting Model II}

In the latter model, the dark energy dominated solution is the only
attractor solution in classical Einstein cosmology, whereas there
exists no attractor in LQC. Numerical results for the interacting
model II are presented in Figs.9-14. Since observations constrain
the interaction to be sub-dominant today, which indicates
$|\Gamma|\ll H_{0}$, we select the parameter $\beta$ to be very
small in the following numerical analysis.

In Fig.9, we plot three-dimensional phase space trajectories of the
universe with $w_{d}=-1.2$ in the stable region in classical
Einstein cosmology. From the figure, we see that the trajectory
curves from different initial conditions are converged at a point,
the position of which is independent of any parameters.

In Fig.10, the evolution of the total EoS $w$ is plotted. It is easy
to see that in the final state $w$ tends to a constant, which equals
to $w_{d}$. It is worth noting that the evolution trajectory of $w$
is not only independent of the theory describing our universe but
also independent of the coupling constant $\beta$. In other words,
the coupling constant do not affect the evolution result.

Fig.11 show the trajectories of scalar factor $a$ versus time for
different values of parameters in LQC. We also take $\kappa^{2}=1$
and $\rho_{c}=1$. The evolution trajectories for different values of
parameters are distinct. The bounce in scale factor occurs later for
greater value of $w_{d}$ or $\beta$. Our universe finally enters an
oscillating phase in LQC.

In Figs.12-14, we plot the evolution trajectories of the Hubble
parameter $H$ and energy density $\rho$ versus time. The parameters
we selected as those above is in the unstable region in LQC. The
universe goes on bouncing forward and backward. The oscillating
frequencies of $H(t)$ and $\rho(t)$ depend not only upon EoS $w_{d}$
but also upon the coupling constant $\beta$.

\section{Conclusions}

In previous sections, we have studied the cosmological evolution of
two interacting dark energy models in classical Einstein and Loop
Quantum Cosmology. We consider two kinds of interaction term between
dark energy and cold dark matter. Note that observations at the
level of the solar system severely constrain non-gravitational
interactions of baryons. So the baryonic matter solely satisfies the
energy conservation equation. By our analysis, we find that
dynamical results in LQC are different from those in classical
Einstein cosmology for both two kinds of interacting models.

In the interacting model I, namely,
$Q=3H(c_{1}\rho_{m}+c_{2}\rho_{d})$, there are two attractors in
both classical Einstein cosmology and LQC. One is a baryon dominated
solution and the other is an accelerated scaling solution. Since the
same results are obtained in both classical Einstein cosmology and
LQC for the baryon dominated attractor, we only focus on the
accelerated scaling solution. Interestingly, we find that if
$w_{d}>-1$, the stable region in LQC is the same as that in
classical Einstein cosmology, while when $w_{d}<-1$, the stable
region in LQC is smaller than that in classical Einstein cosmology.
The total EoS $w$ approaches finally to a constant, which depends on
EoS $w_{d}$ and the coupling constants $c_{1}$ and $c_{2}$, but is
independent of the theory describing our universe. When we select
the parameters in the unstable region in LQC, the universe
experience bouncing, which can resolve the singularity problem. The
bounce in scale factor occurs later for greater value of $w_{d}$,
$c_{1}$ or $c_{2}$. Furthermore, the oscillating frequencies are
distinct for different parameters.

However, in the interacting model II with $Q=\Gamma\rho_{m}$, there
exists one attractor in classical Einstein cosmology for $w_{d}<-1$,
which is a dark energy dominated solution rather than a scaling
solution, whereas in LQC, all fixed points is not stable. Thus,
there exists no scaling solutions in the interacting model II. So
this kind of interacting dark energy model can not be regarded as a
candidate to alleviate the coincidence problem in both classical
Einstein and loop quantum cosmology. In classical Einstein
cosmology, the final state $w_{*}$ is a constant, which equals to
$w_{d}$ and is independent of the coupling constant $\beta$. The
bounce in scale factor occurs later for greater value of $w_{d}$ or
$\beta$. Our universe finally enters an oscillating phase in LQC.
Moreover, the oscillating frequencies are significantly different
for varied parameters.

In summary, the interacting model I may alleviate the coincidence
problem in both classical Einstein and loop quantum cosmology,
depending on the values of the parameters selected in the model.
However, the interacting model II can not be regarded as a candidate
to alleviate the coincidence problem in both kinds of cosmology.
Thus, dynamical results are different not only in different theories
describing the universe but also in different interacting models. In
addition, the results that our universe finally enters an
oscillating phase in LQC, which are different from the those
obtained in classical Einstein cosmology, show that LQC allow us the
possibility of resolving future singularities. Therefore, the
quantum gravity effect may be manifested in large scale in the
interacting dark energy models.

\section*{Acknowledgements}

This work is a part of projects 10675019 and 10975017 supported by
NSFC.

\newpage

\begin{figure}[tbp]
\begin{center}
\includegraphics[width=0.3\textwidth]{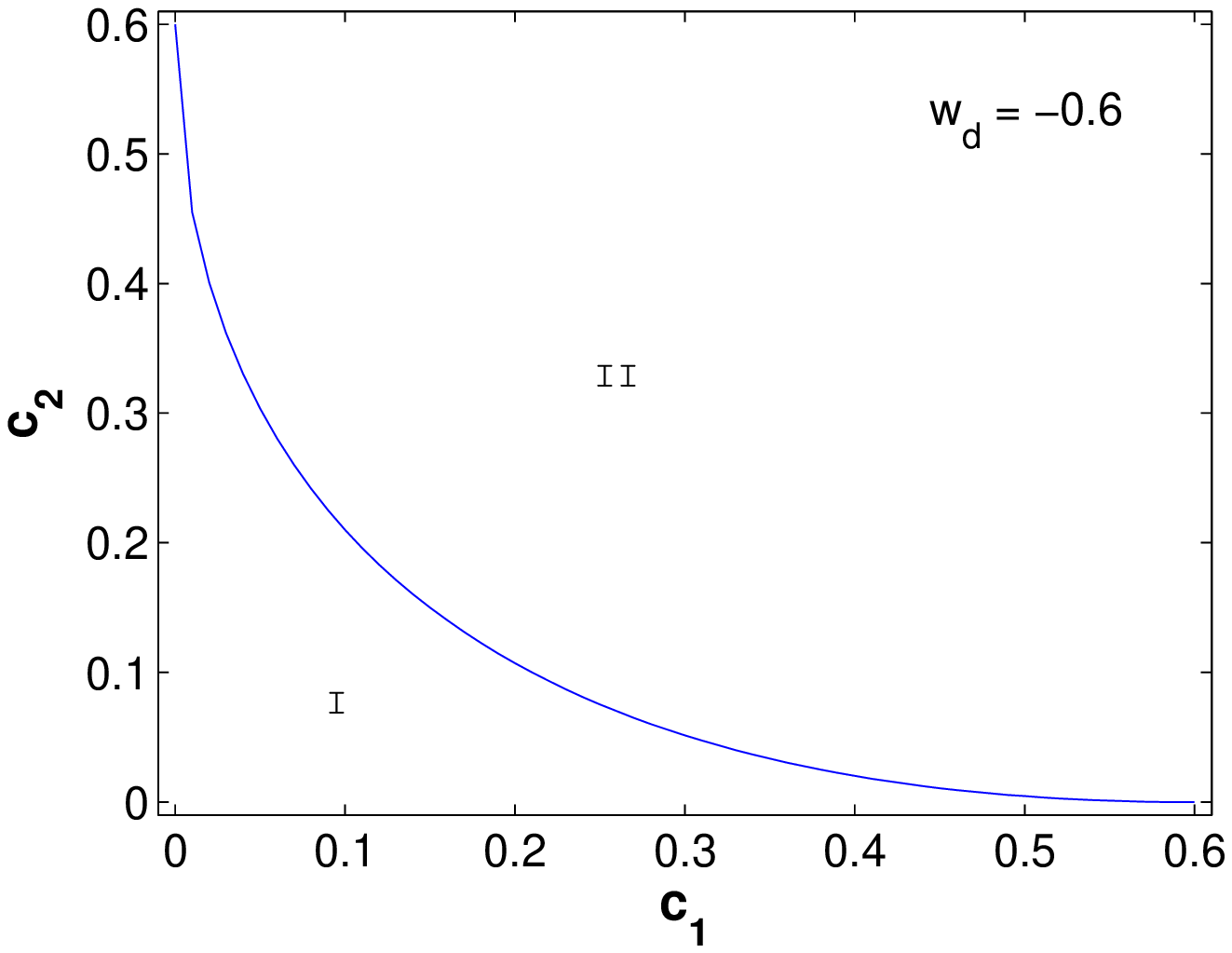}%
\includegraphics[width=0.3\textwidth]{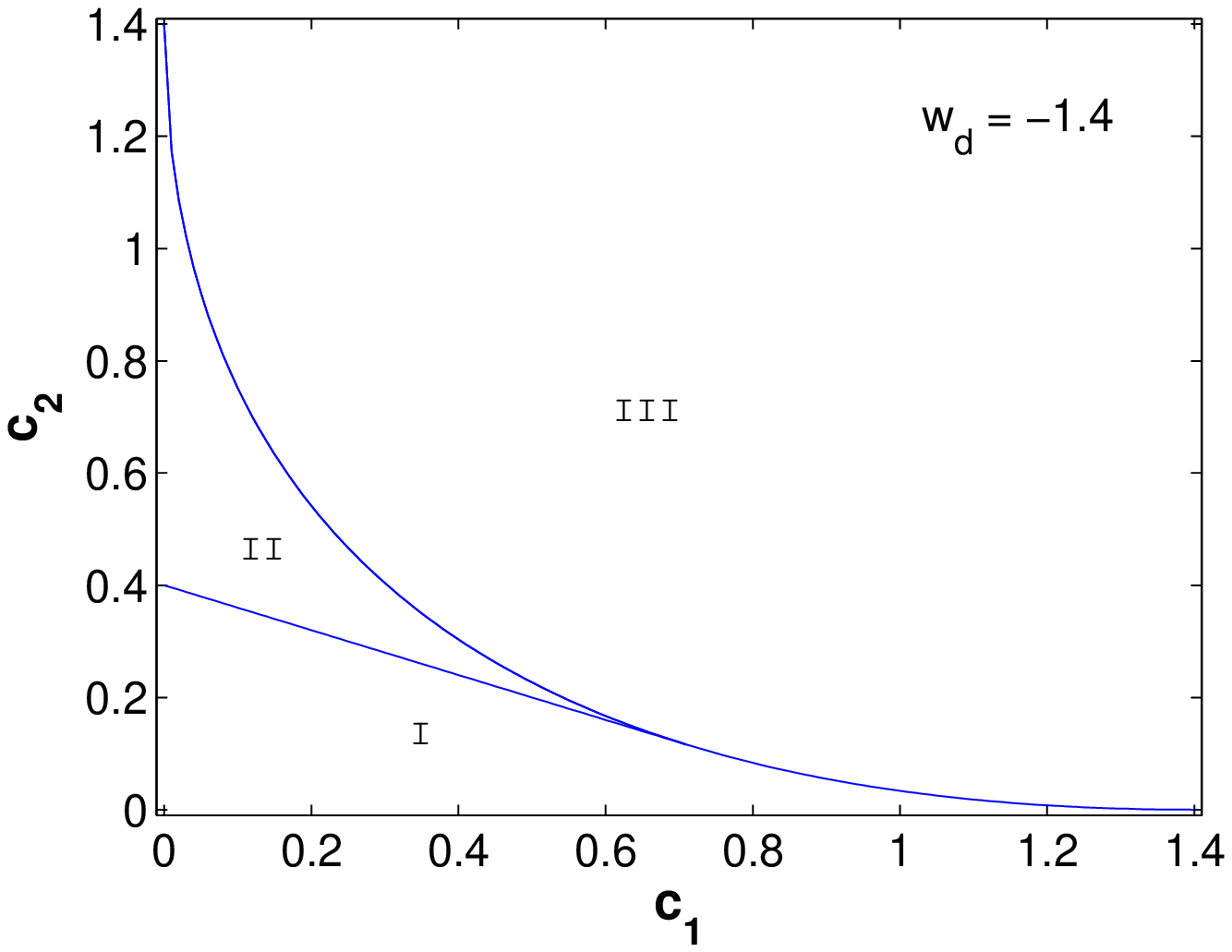}
\end{center}
\caption{The stable regions in the ($c_{1}$,$c_{2}$) parameter space
with the fixed $w_{d}$ for the interacting model I. In the left plot
($w_{d}>-1$), the critical point B is stable in the region I in
classical Einstein cosmology and LQC. In the right plot
($w_{d}<-1$), point B is stable in the regions I+II in classical
Einstein cosmology, whereas in LQC, point B is an attractor only in
the region II. The region II in the left and the region III in the
right represent the regions of the physically meaningless solution.}
\label{Fig.1}
\end{figure}

\begin{figure}[tbp]
\begin{center}
\includegraphics[width=0.3\textwidth]{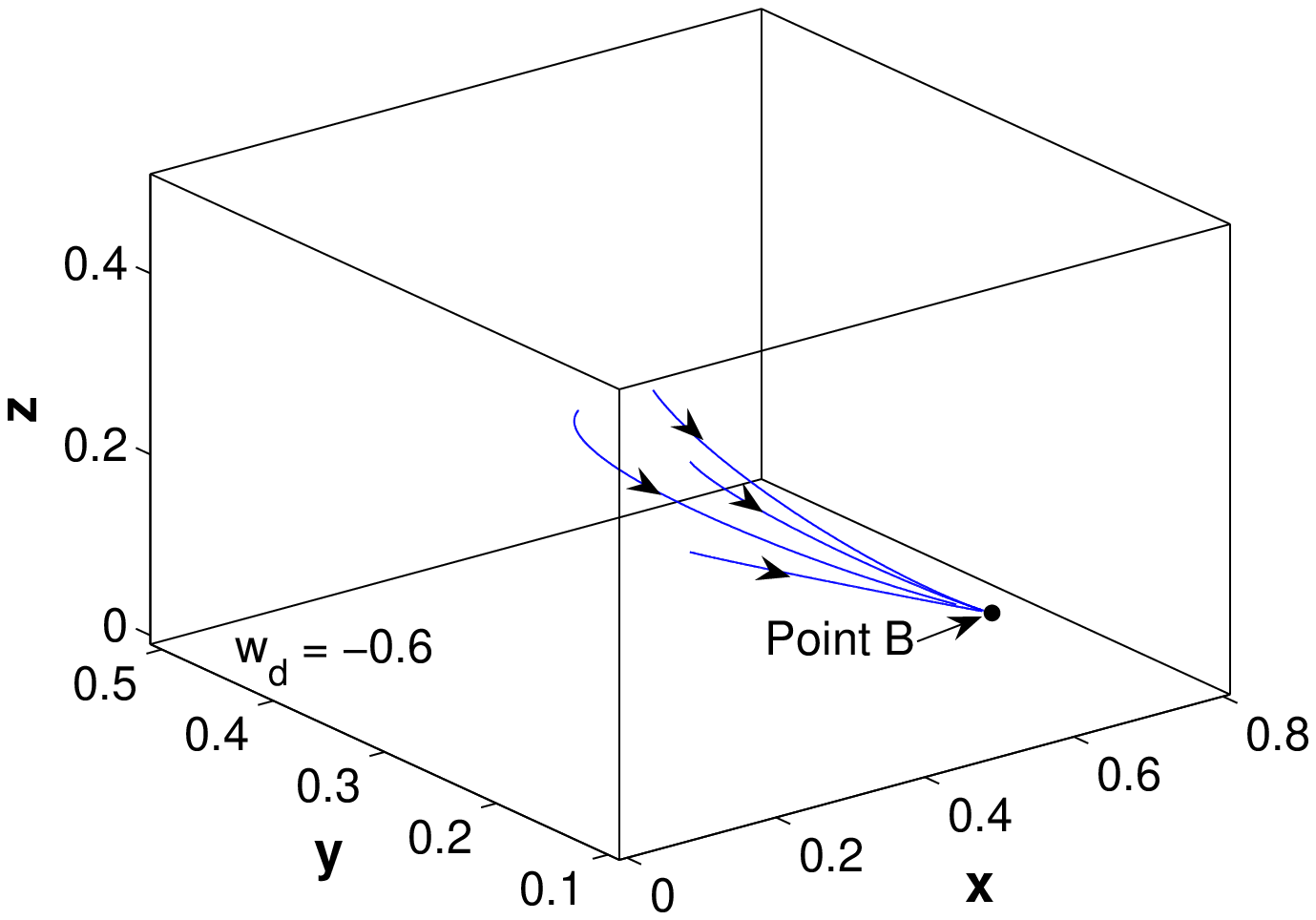}%
\includegraphics[width=0.3\textwidth]{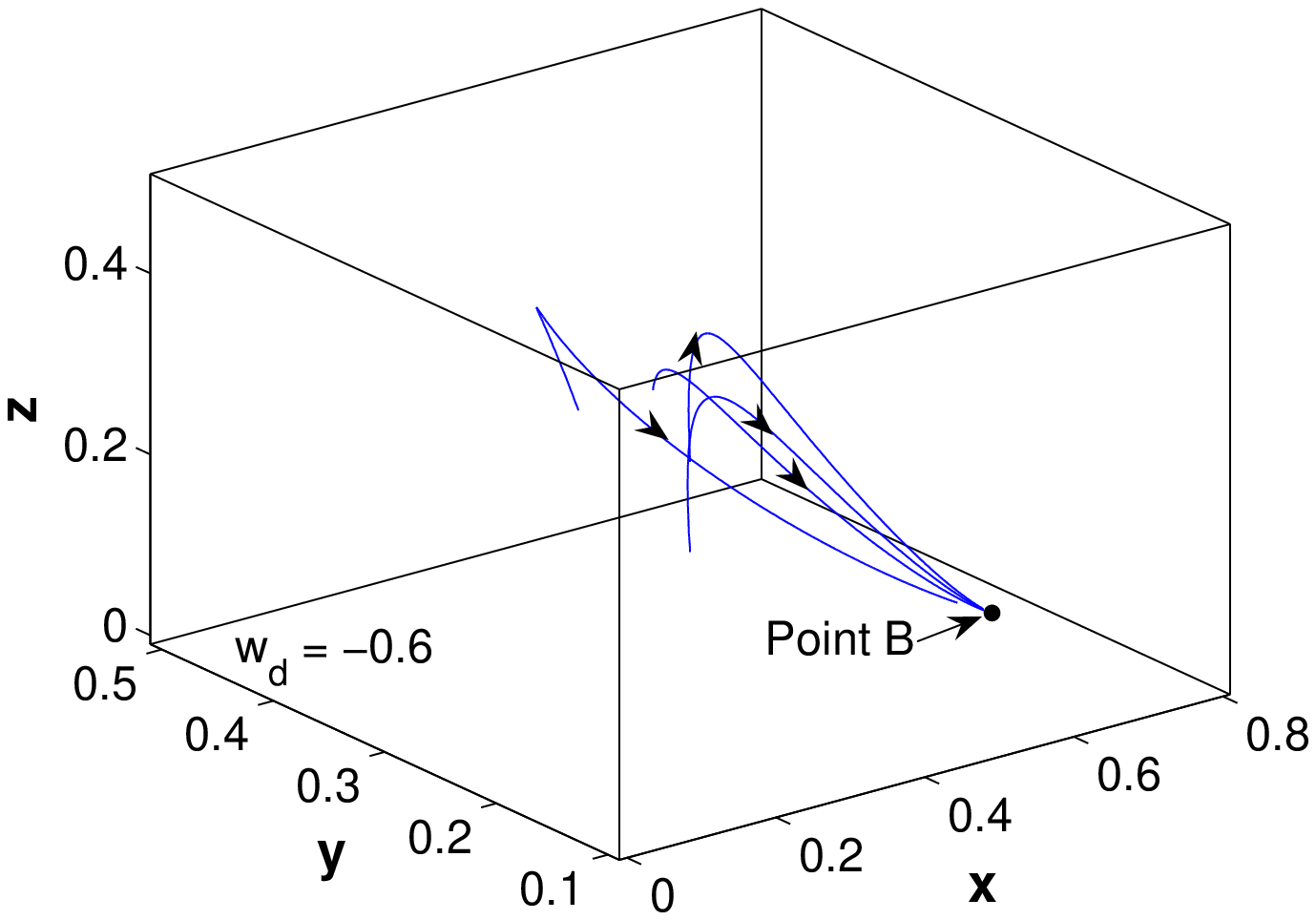}\\
\includegraphics[width=0.3\textwidth]{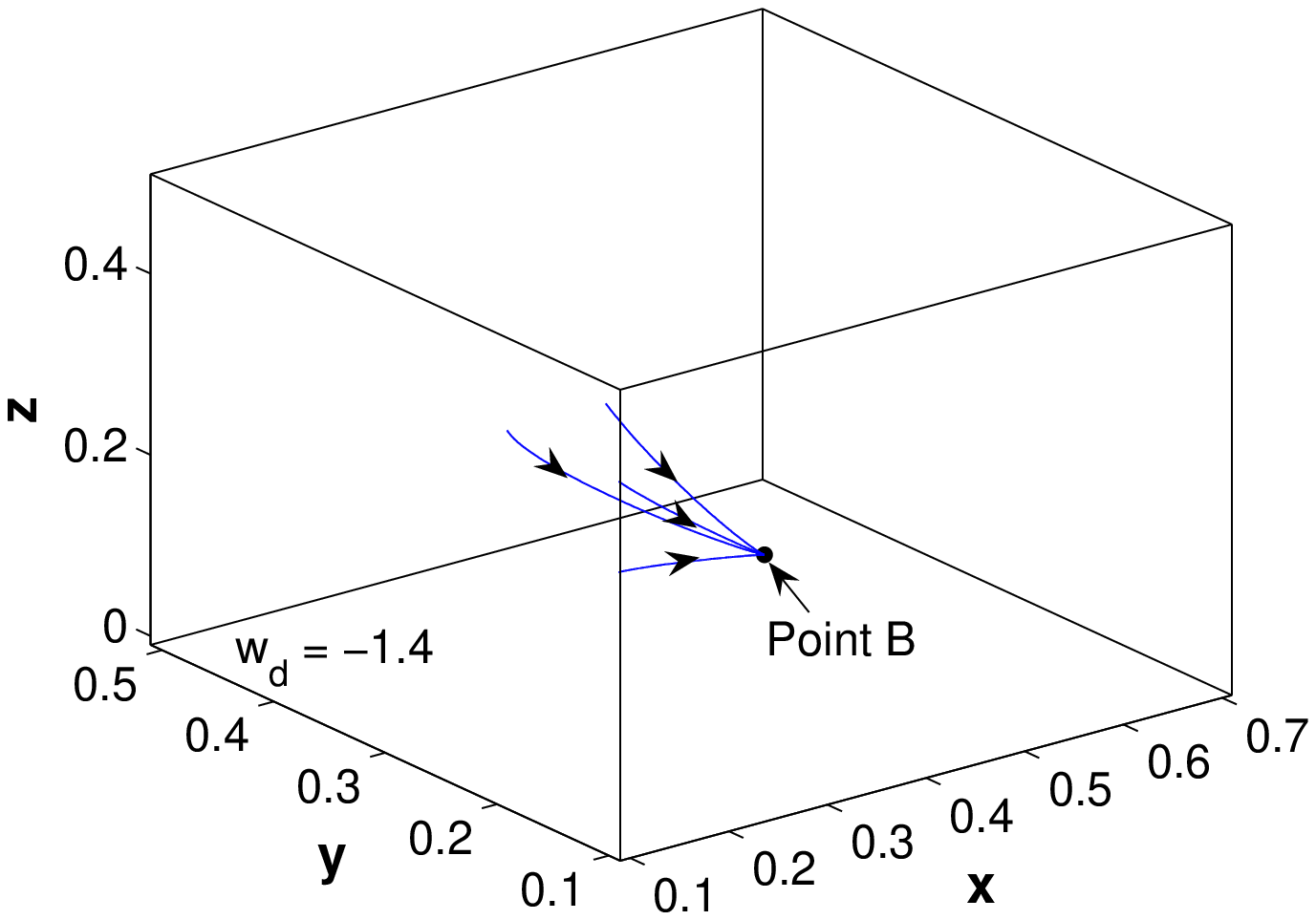}%
\includegraphics[width=0.3\textwidth]{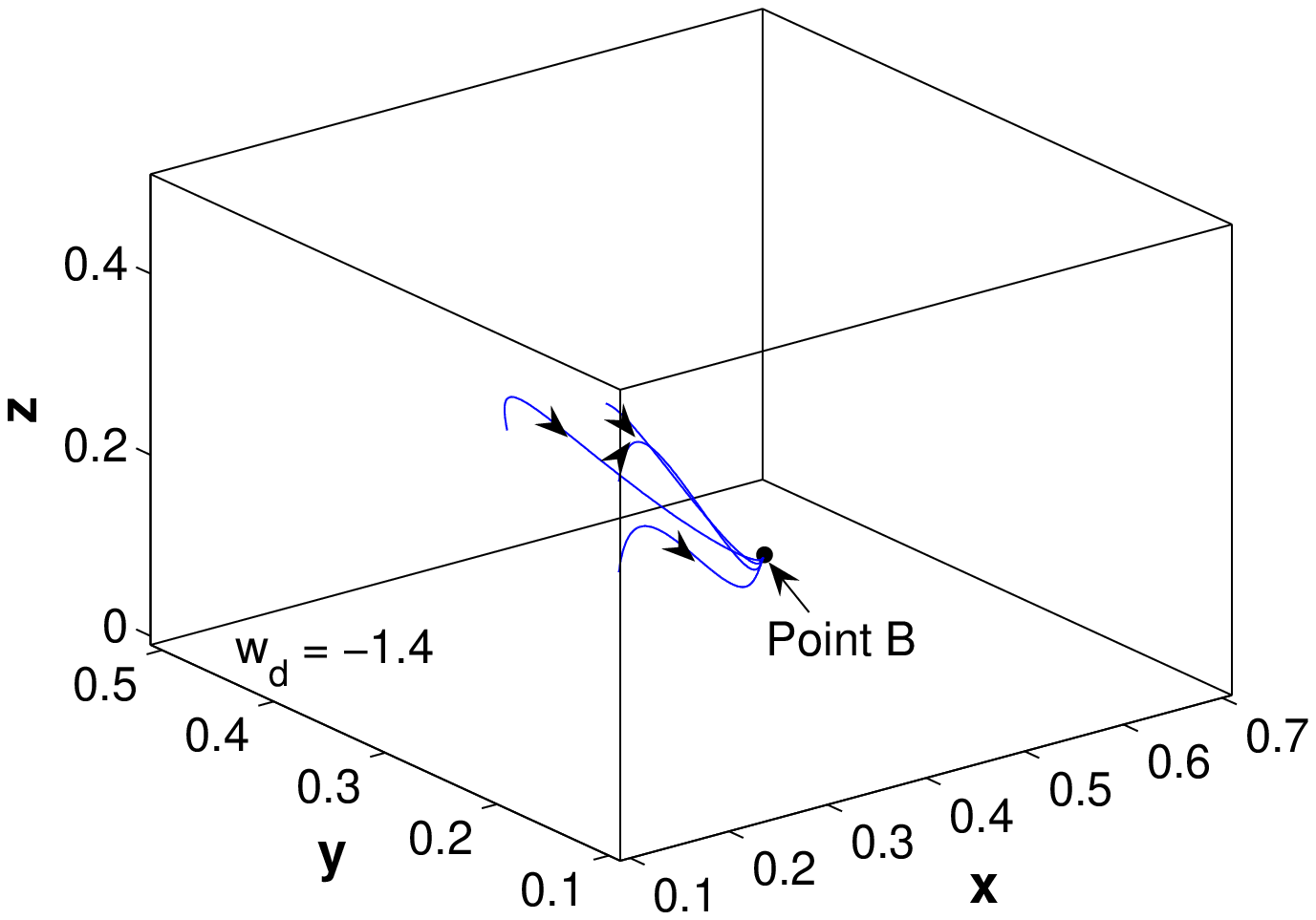}
\end{center}
\caption{Three-dimensional phase space of $(x,y,z)$ with the fixed
$w_{d}$ for the interacting model I. The left and right two plots
respectively denote classical Eintein cosmology and LQC. In the top
two plots, we select the parameters $c_{1}=0.1$ and $c_{2}=0.15$.
The bottom two plots is for $c_{1}=0.1$ and $c_{2}=0.5$.}
\label{Fig.1}
\end{figure}

\begin{figure}[tbp]
\begin{center}
\includegraphics[width=0.3\textwidth]{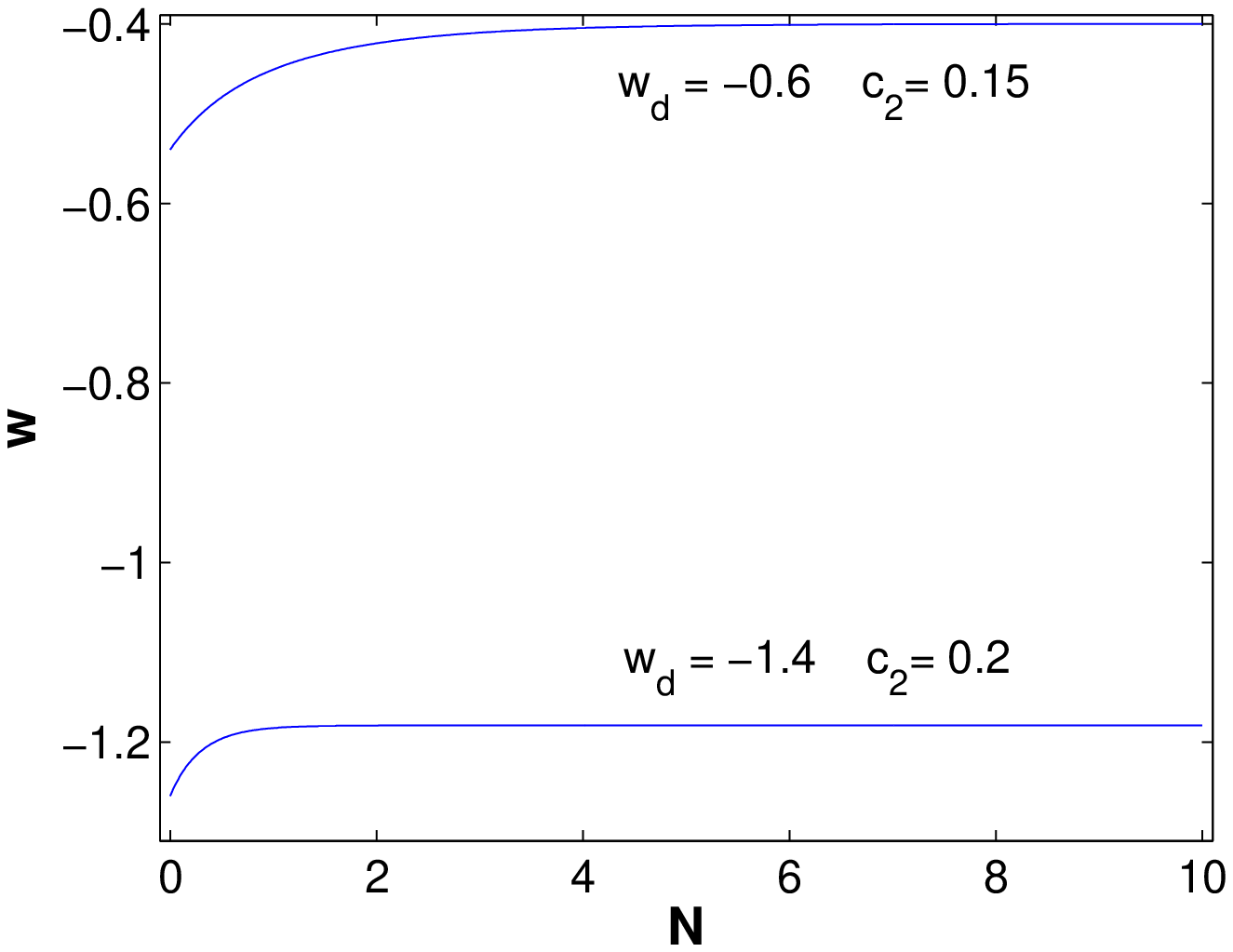}%
\includegraphics[width=0.3\textwidth]{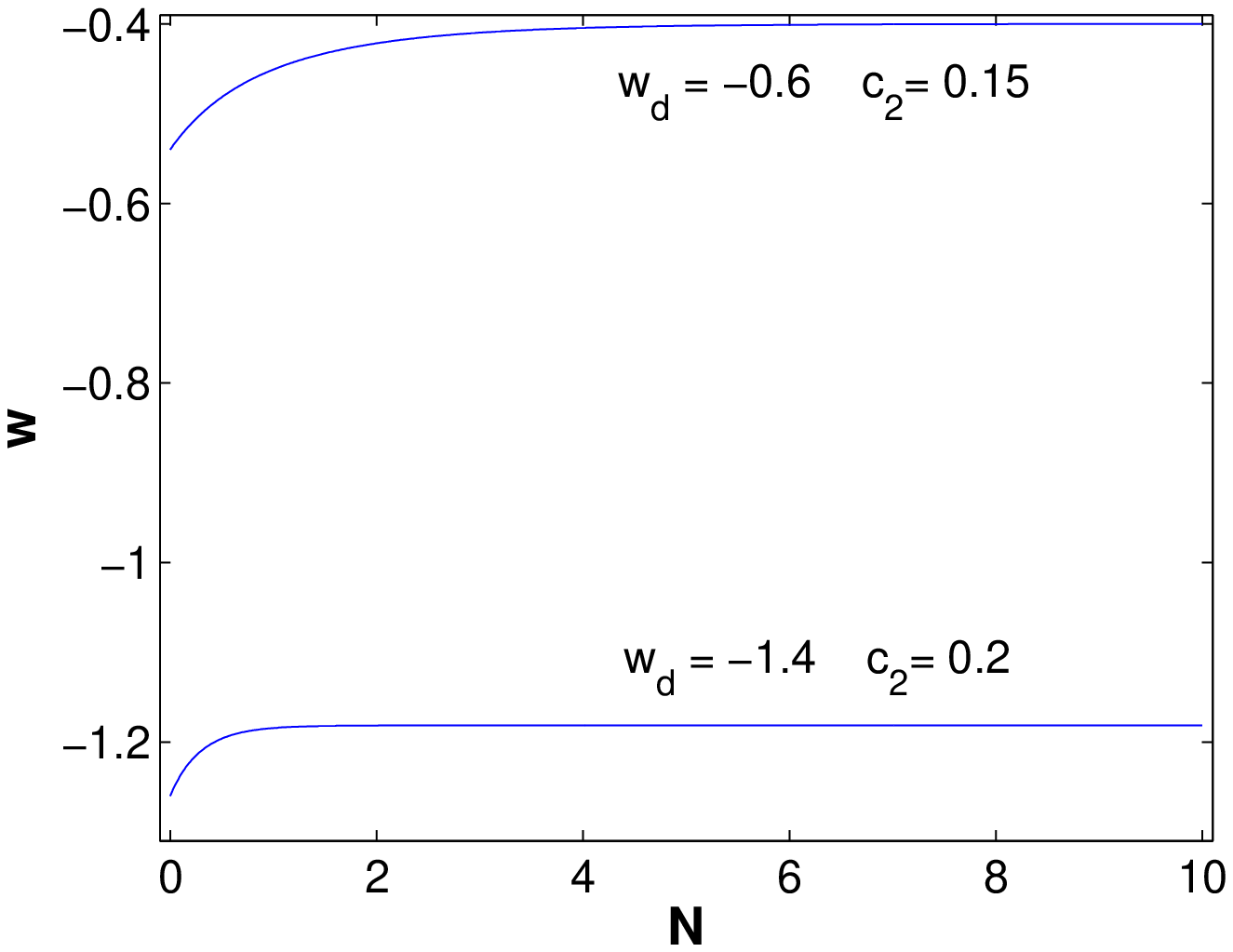}
\end{center}
\caption{The evolution of the EoS of total cosmic fluid $w$ with the
fixed $w_{d}$ for the interacting model I. The left is for classical
Einstein cosmology and the right is for LQC. The parameter $c_{1}$
is chosen as 0.1.} \label{Fig.1}
\end{figure}

\begin{figure}[tbp]
\begin{center}
\includegraphics[width=0.3\textwidth]{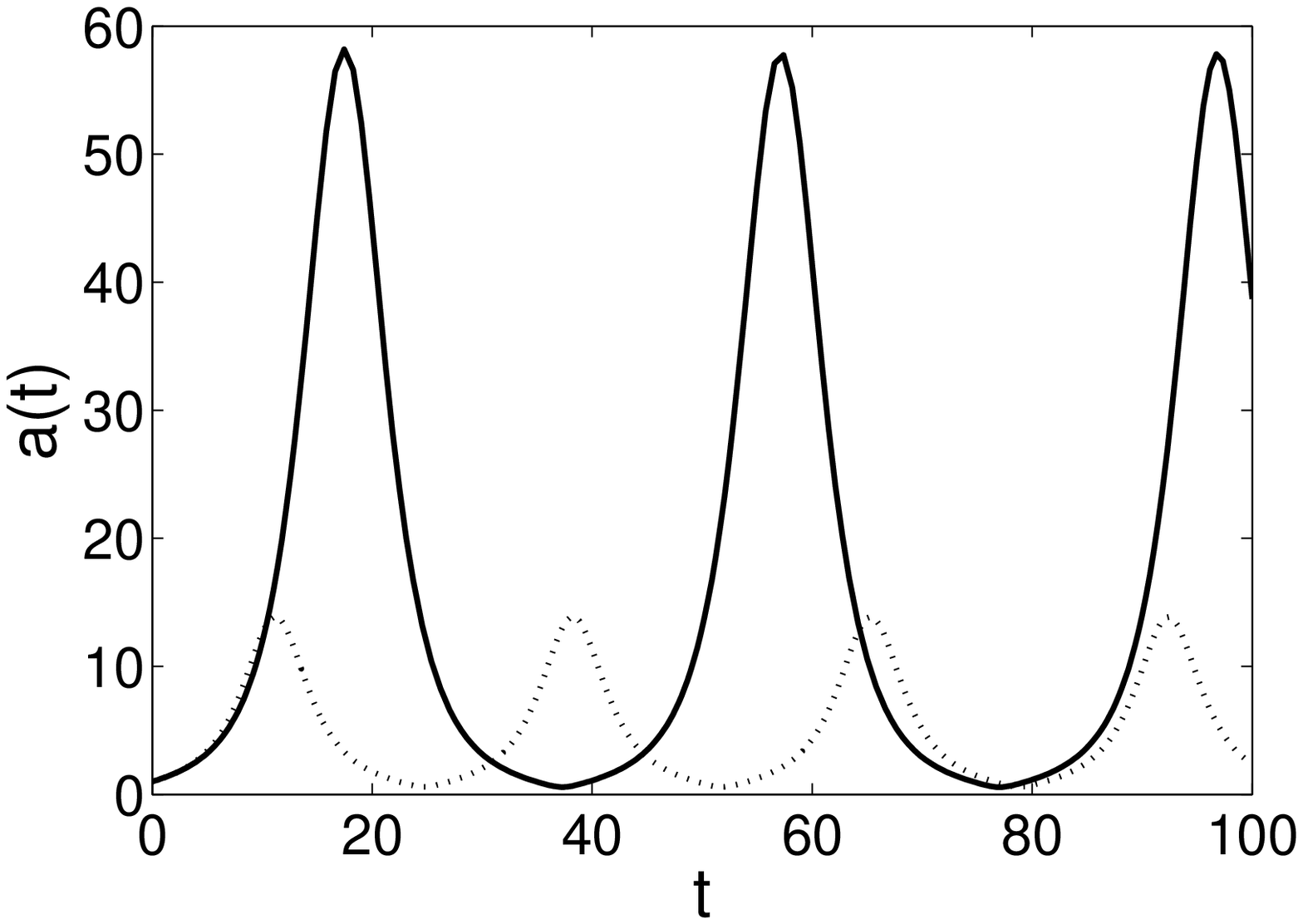}%
\includegraphics[width=0.3\textwidth]{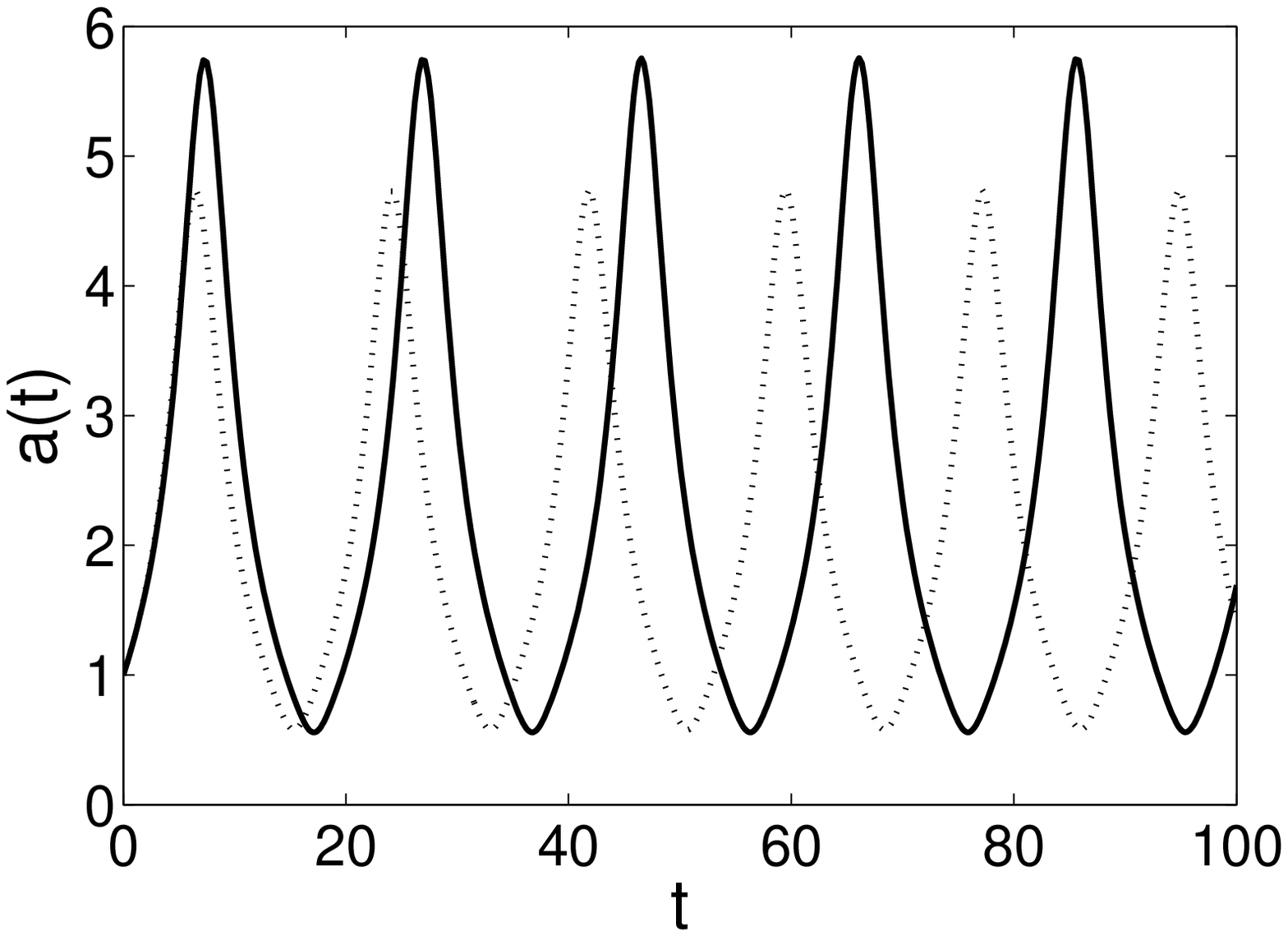}%
\includegraphics[width=0.3\textwidth]{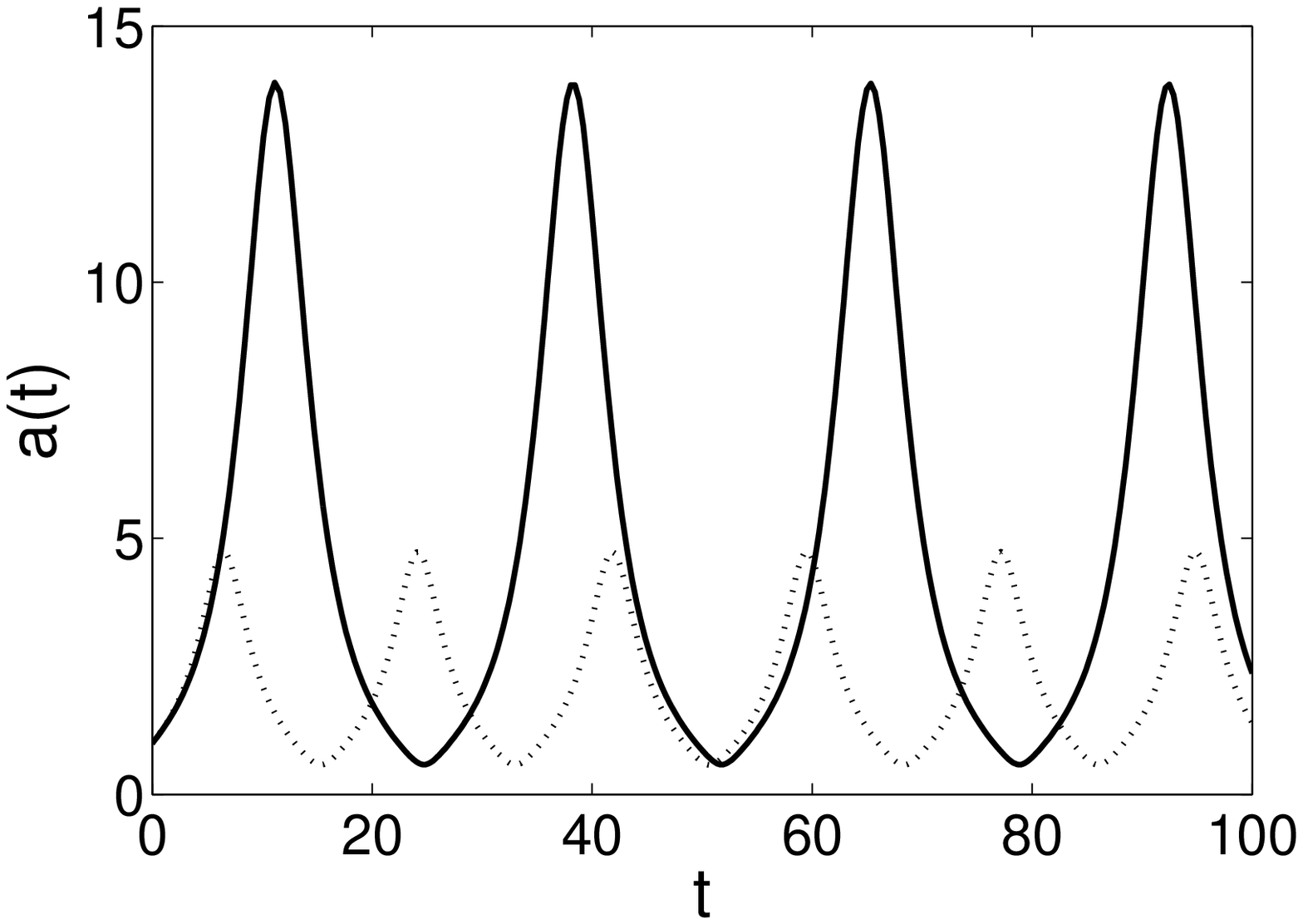}
\end{center}
\caption{The plots of scalar factor $a$ as a function of time for
the fixed parameters. The left plot is for $w_{d}=-1.4$ and
$c_{1}=0.1$. The solid and dotted lines respectively correspond to
$c_{2}=0.2$ and $c_{2}=0.1$. The middle plot is for $w_{d}=-1.6$ and
$c_{2}=0.1$. The solid and dotted lines respectively correspond to
$c_{1}=0.3$ and $c_{1}=0.1$. The right plot is for
$c_{1}=c_{2}=0.1$. The solid and dotted lines respectively
correspond to $w_{d}=-1.4$ and $w_{d}=-1.6$.} \label{Fig.1}
\end{figure}

\begin{figure}[tbp]
\begin{center}
\includegraphics[width=0.3\textwidth]{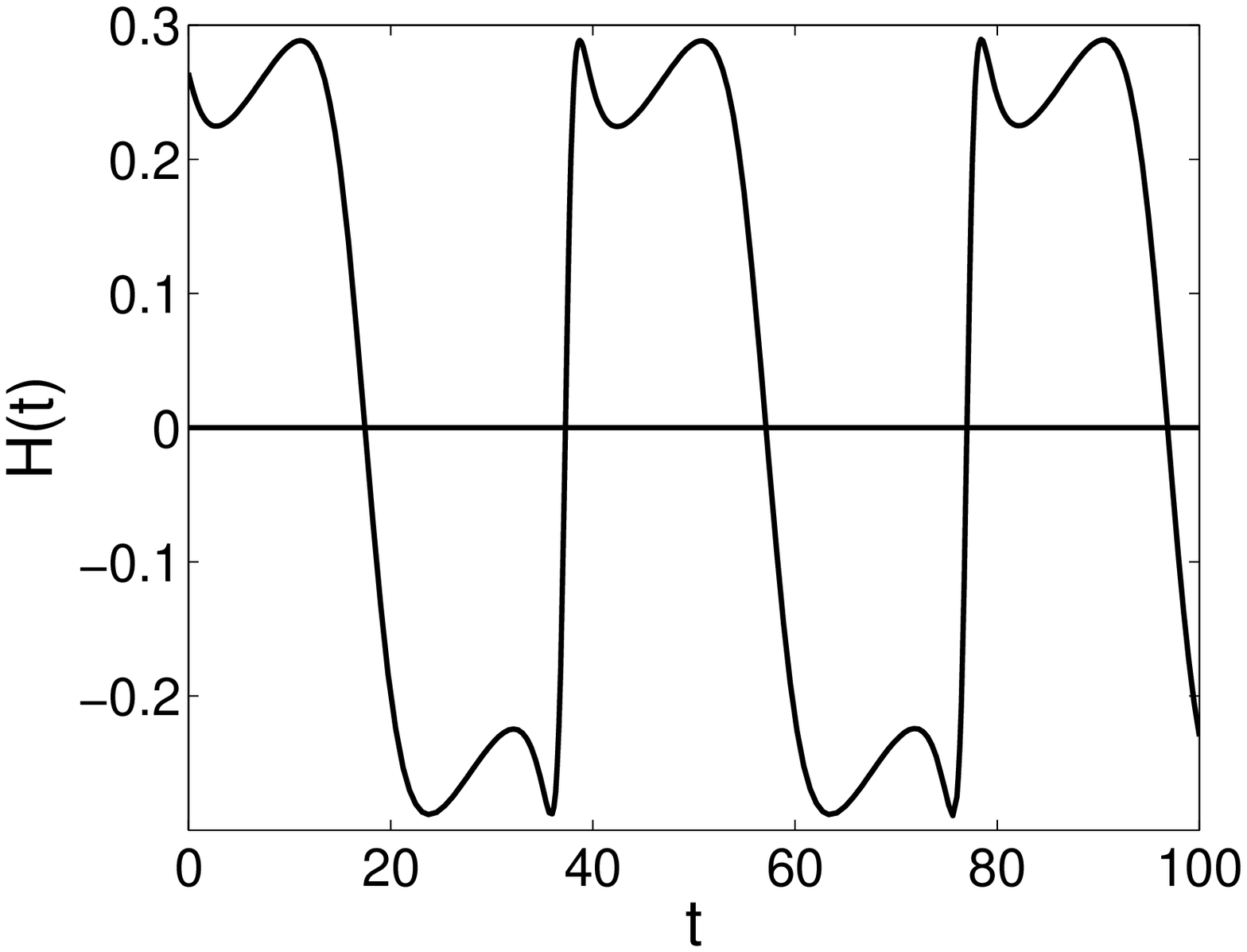}%
\includegraphics[width=0.3\textwidth]{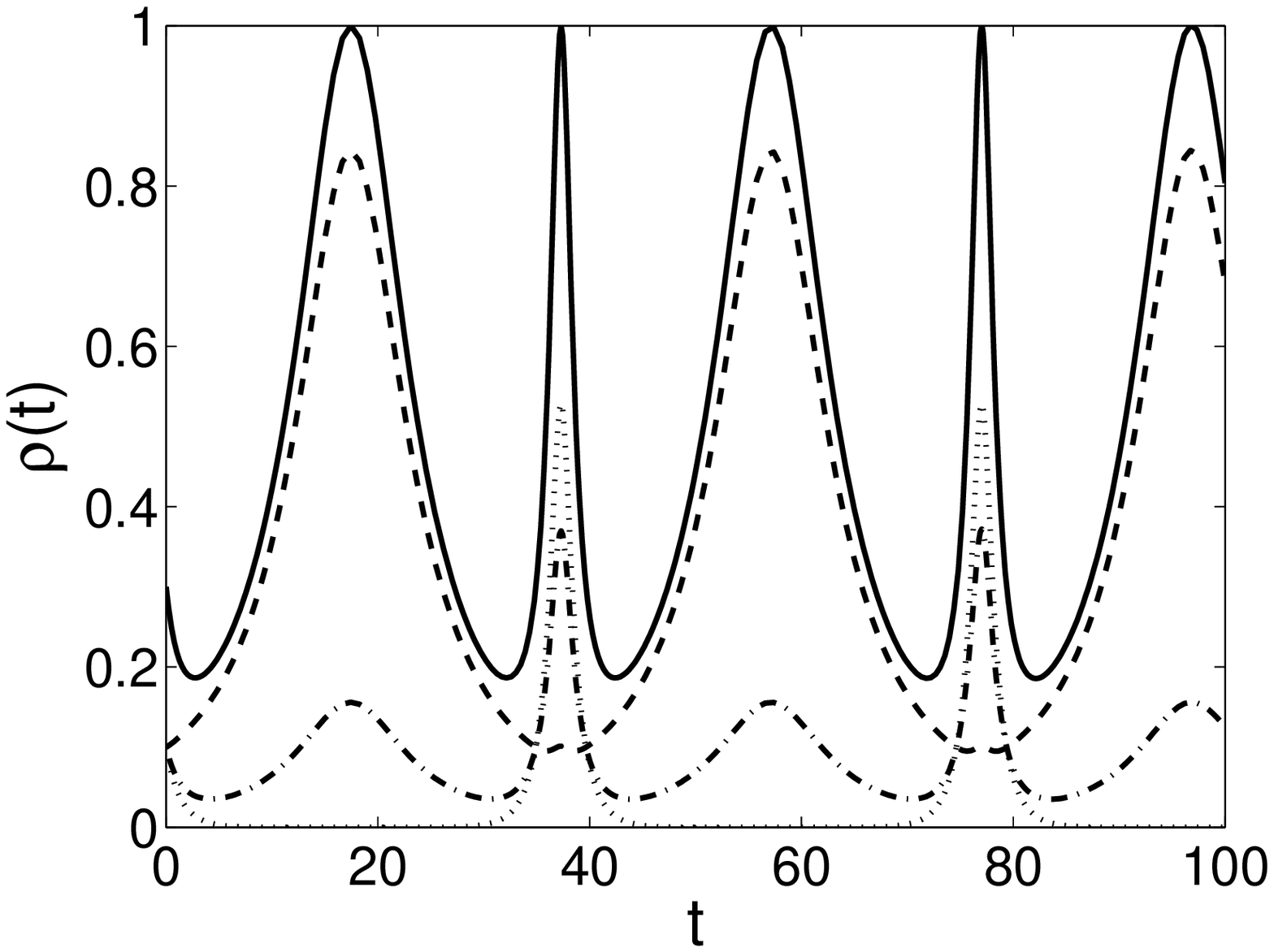}
\end{center}
\caption{The evolution of the Hubble parameter $H$ and the energy
density $\rho$ with respect to time with
$w_{d}=-1.4,c_{1}=0.1,c_{2}=0.2$. The solid, dashed, dash-dotted and
dotted lines correspond to $\rho$, $\rho_{d}$, $\rho_{m}$ and
$\rho_{b}$, respectively. } \label{Fig.1}
\end{figure}

\begin{figure}[tbp]
\begin{center}
\includegraphics[width=0.3\textwidth]{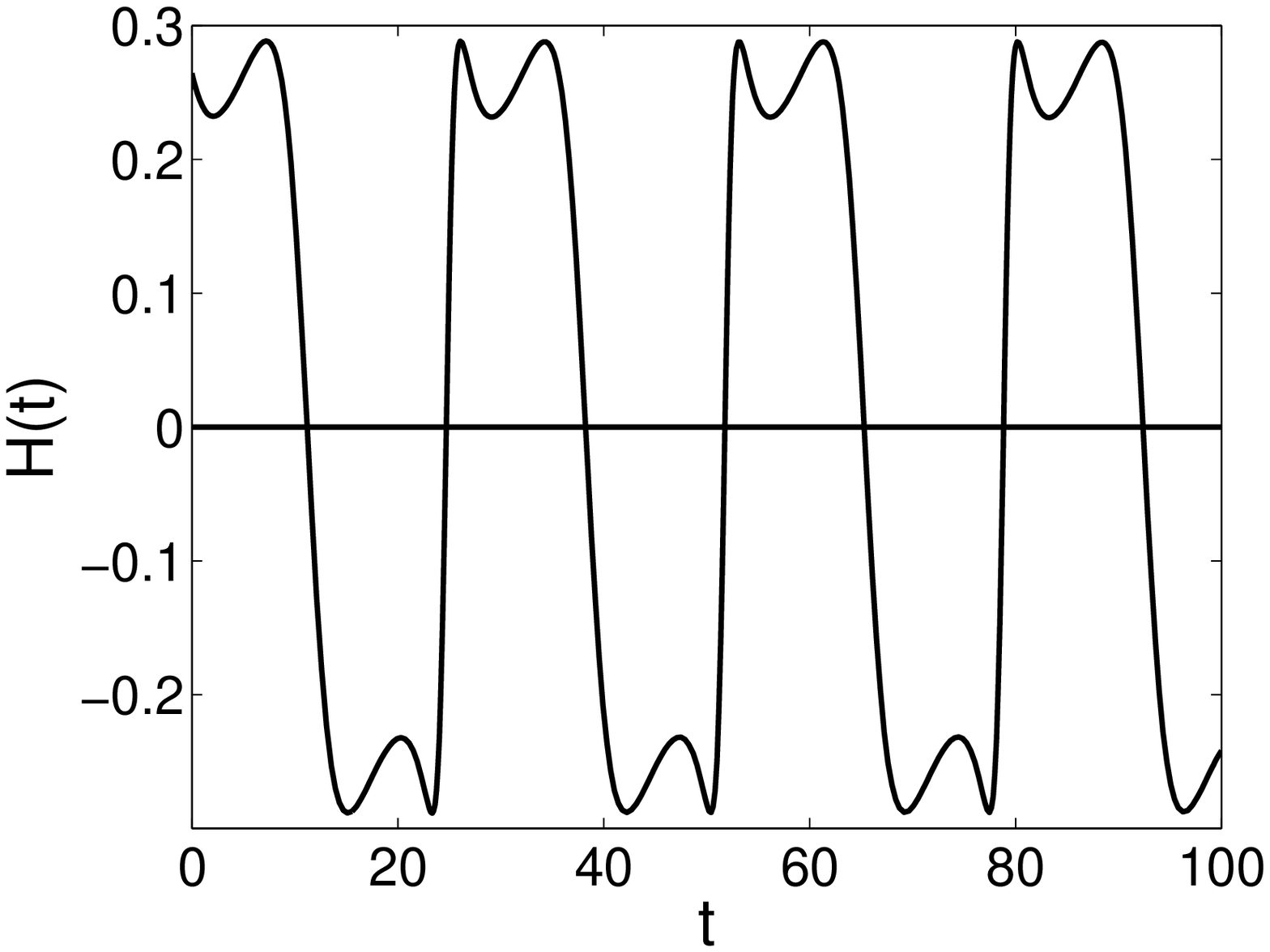}%
\includegraphics[width=0.3\textwidth]{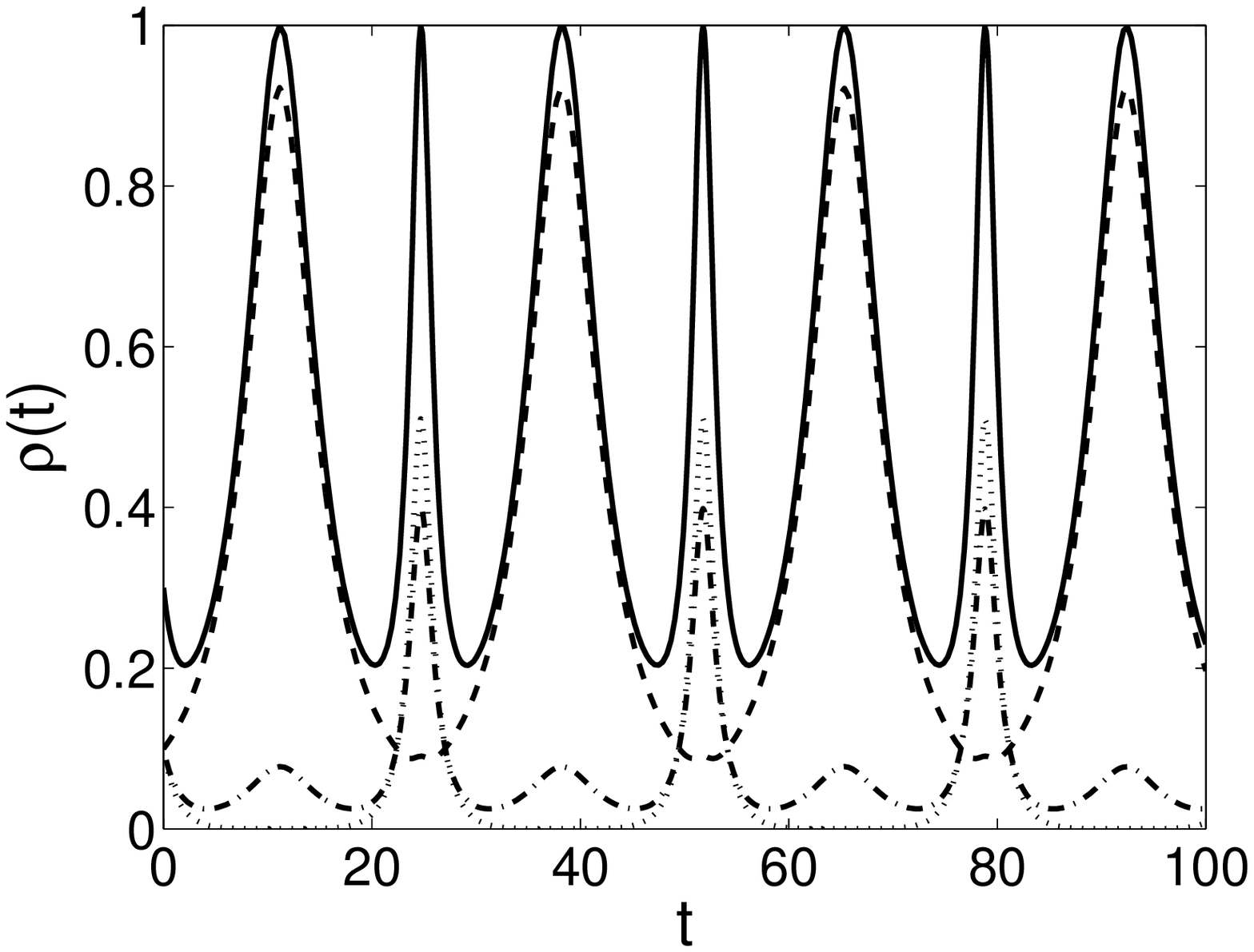}
\end{center}
\caption{The evolution of the Hubble parameter $H$ and the energy
density $\rho$ with respect to time with
$w_{d}=-1.4,c_{1}=0.1,c_{2}=0.1$. The solid, dashed, dash-dotted and
dotted lines correspond to $\rho$, $\rho_{d}$, $\rho_{m}$ and
$\rho_{b}$, respectively. } \label{Fig.1}
\end{figure}

\begin{figure}[tbp]
\begin{center}
\includegraphics[width=0.3\textwidth]{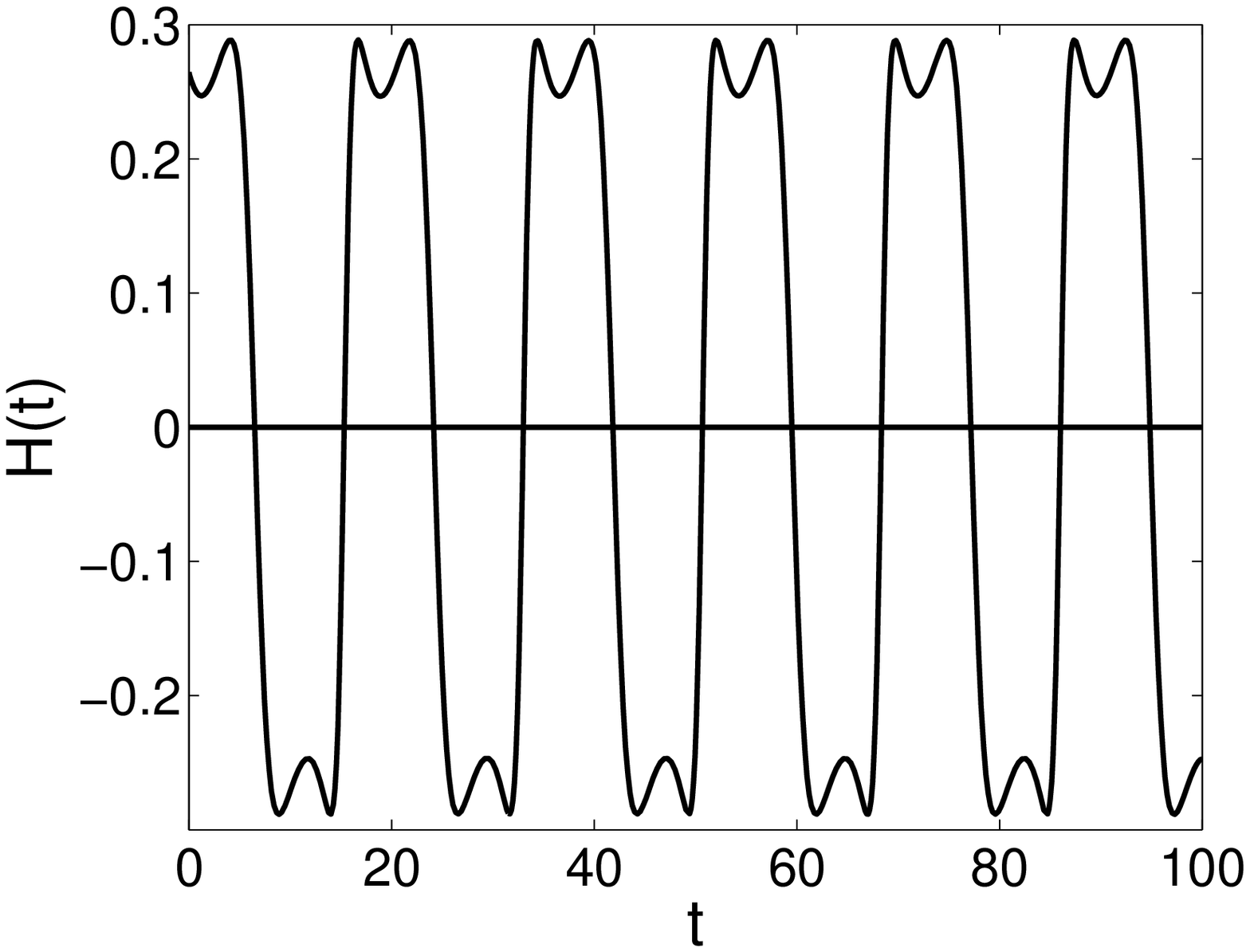}%
\includegraphics[width=0.3\textwidth]{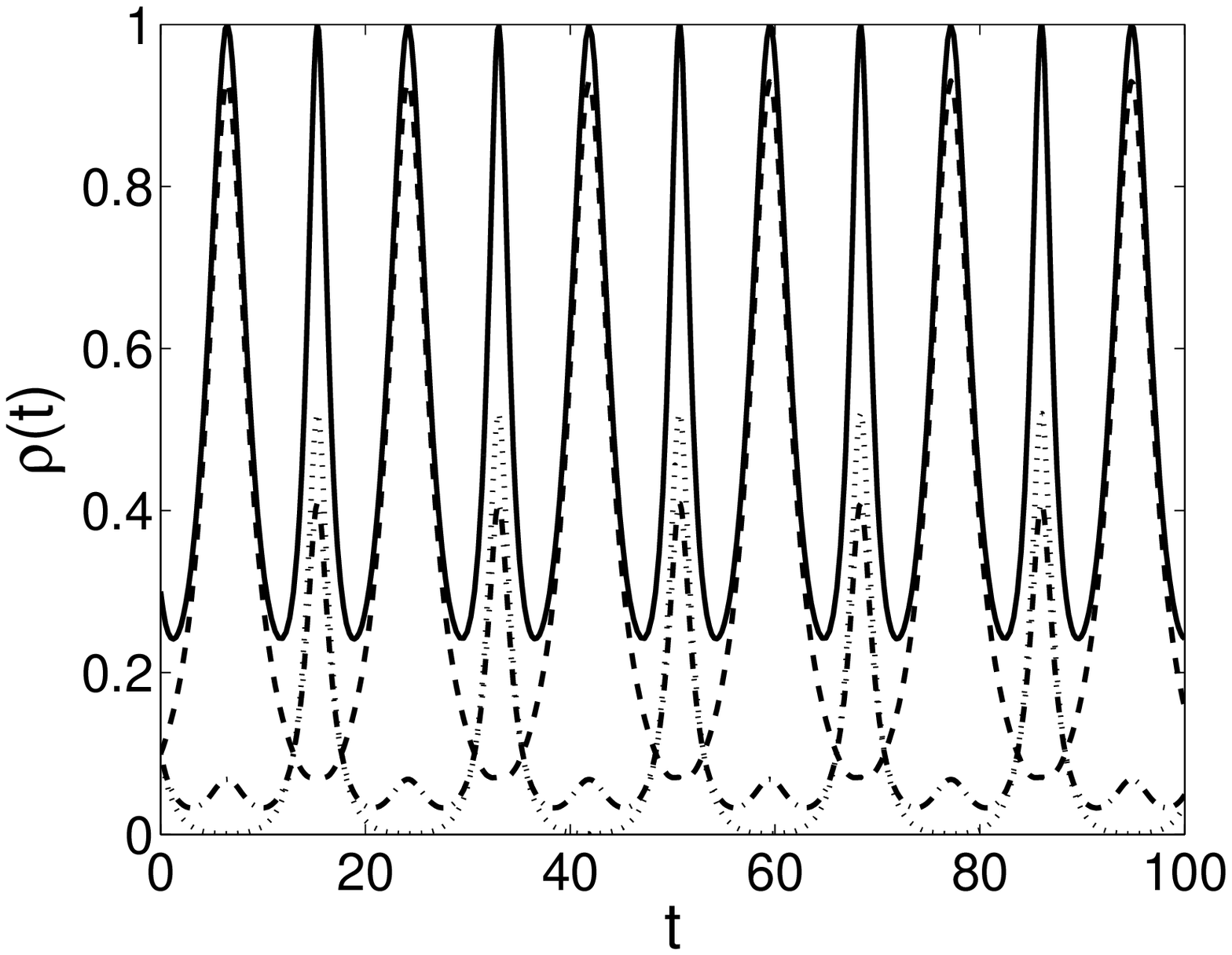}
\end{center}
\caption{The evolution of the Hubble parameter $H$ and the energy
density $\rho$ with respect to time with
$w_{d}=-1.6,c_{1}=0.1,c_{2}=0.1$. The solid, dashed, dash-dotted and
dotted lines correspond to $\rho$, $\rho_{d}$, $\rho_{m}$ and
$\rho_{b}$, respectively. } \label{Fig.1}
\end{figure}

\begin{figure}[tbp]
\begin{center}
\includegraphics[width=0.3\textwidth]{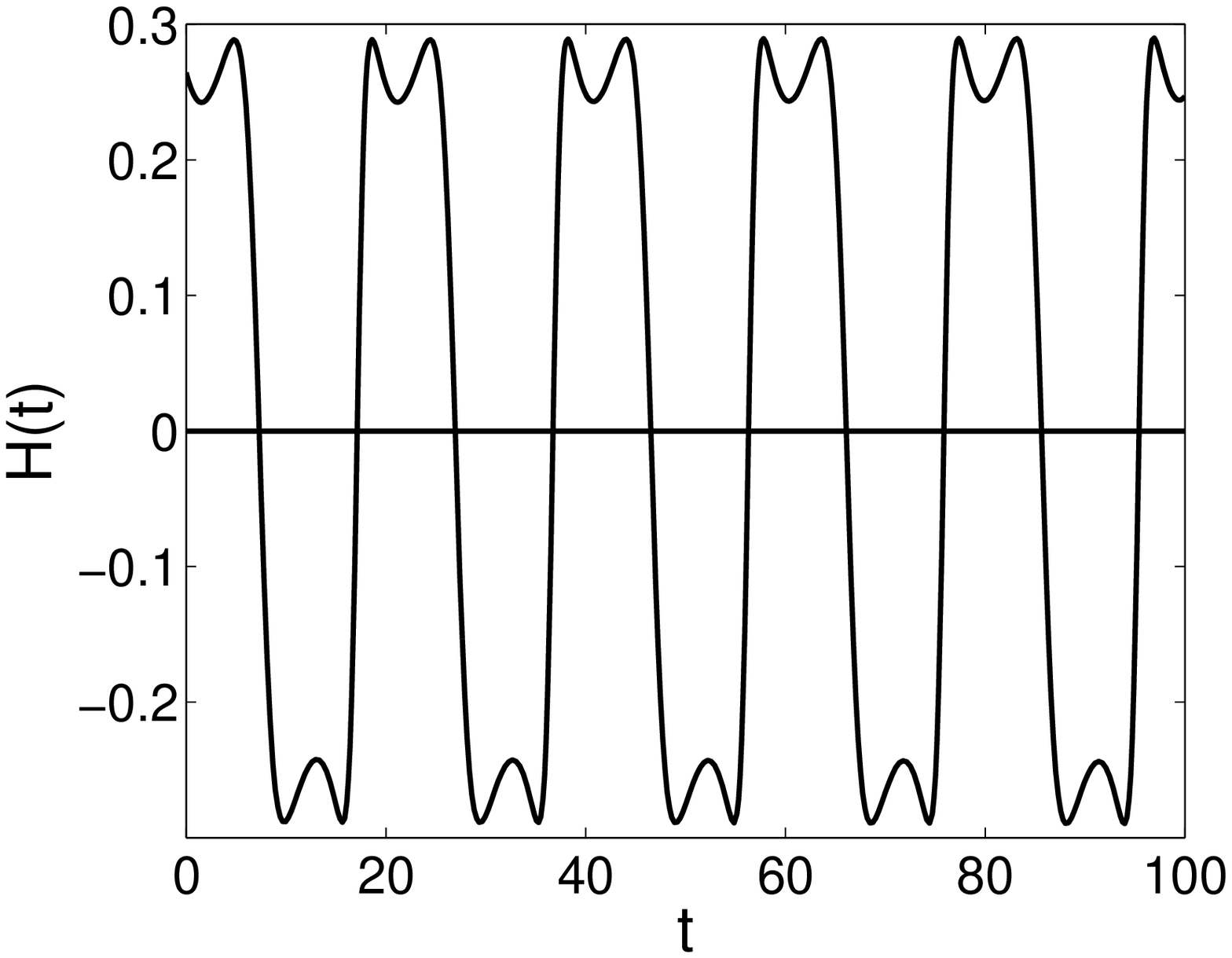}%
\includegraphics[width=0.3\textwidth]{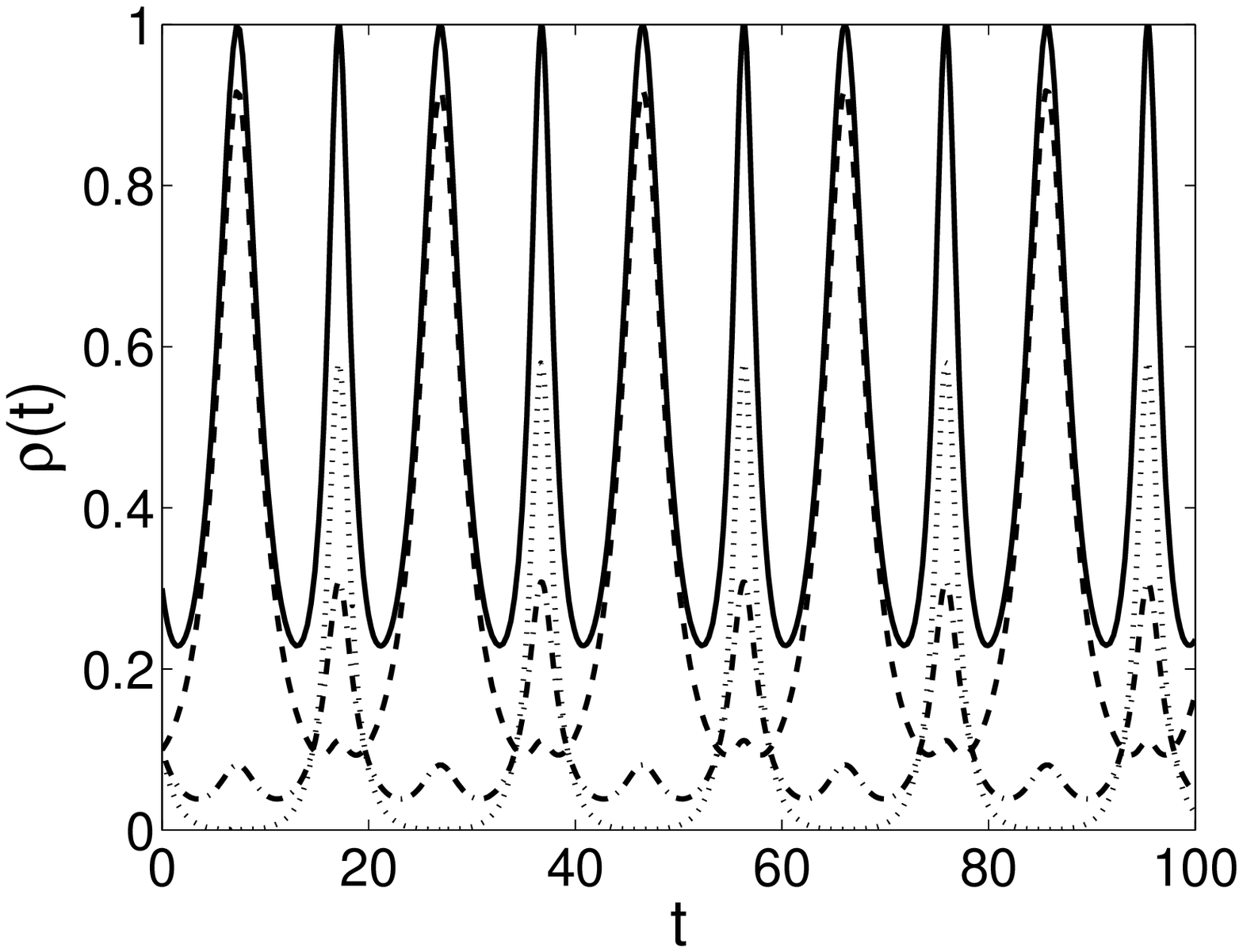}
\end{center}
\caption{The evolution of the Hubble parameter $H$ and the energy
density $\rho$ with respect to time with
$w_{d}=-1.6,c_{1}=0.3,c_{2}=0.1$. The solid, dashed, dash-dotted and
dotted lines correspond to $\rho$, $\rho_{d}$, $\rho_{m}$ and
$\rho_{b}$, respectively. } \label{Fig.1}
\end{figure}

\begin{figure}[tbp]
\begin{center}
\includegraphics[width=0.3\textwidth]{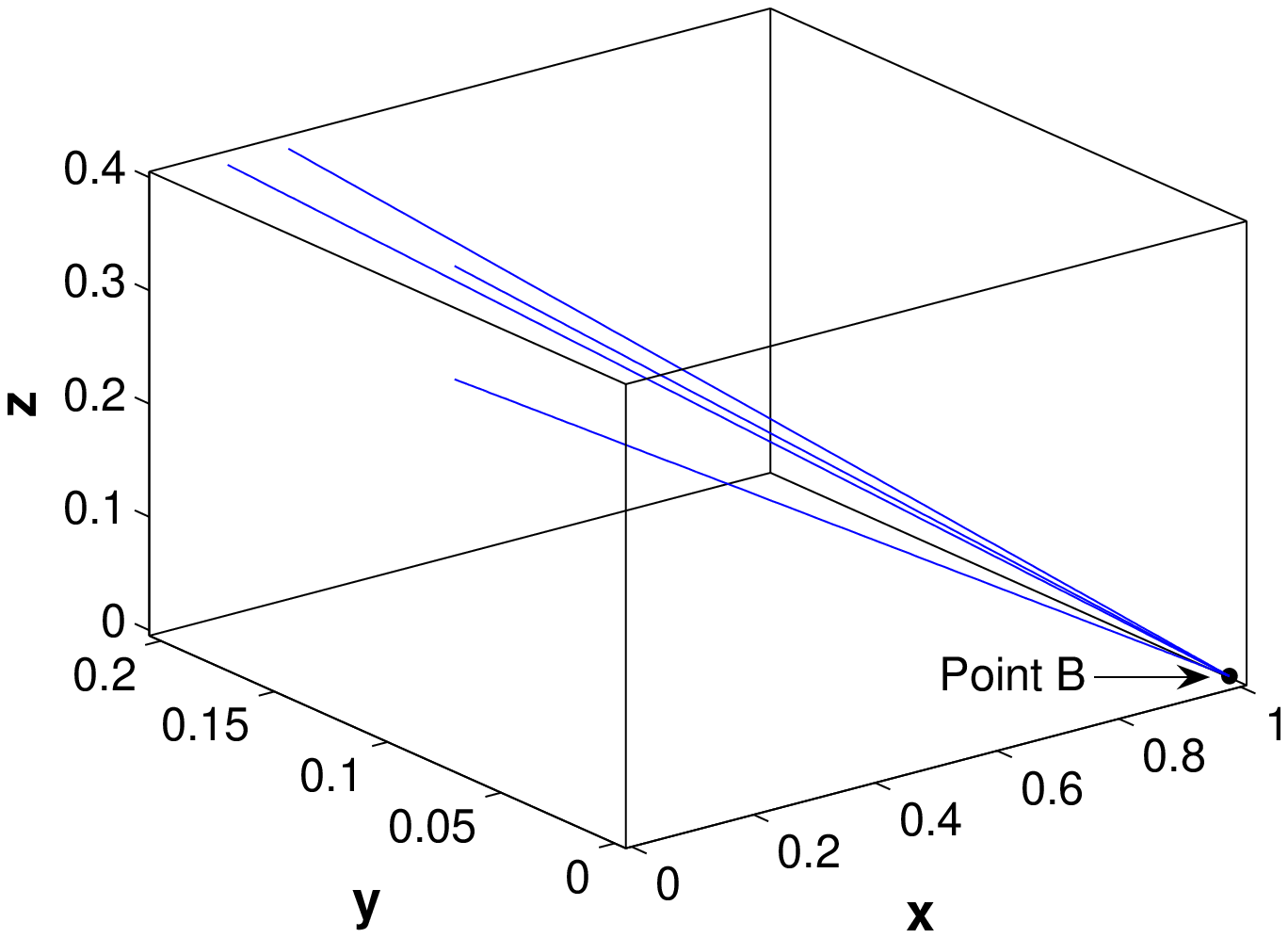}%
\includegraphics[width=0.3\textwidth]{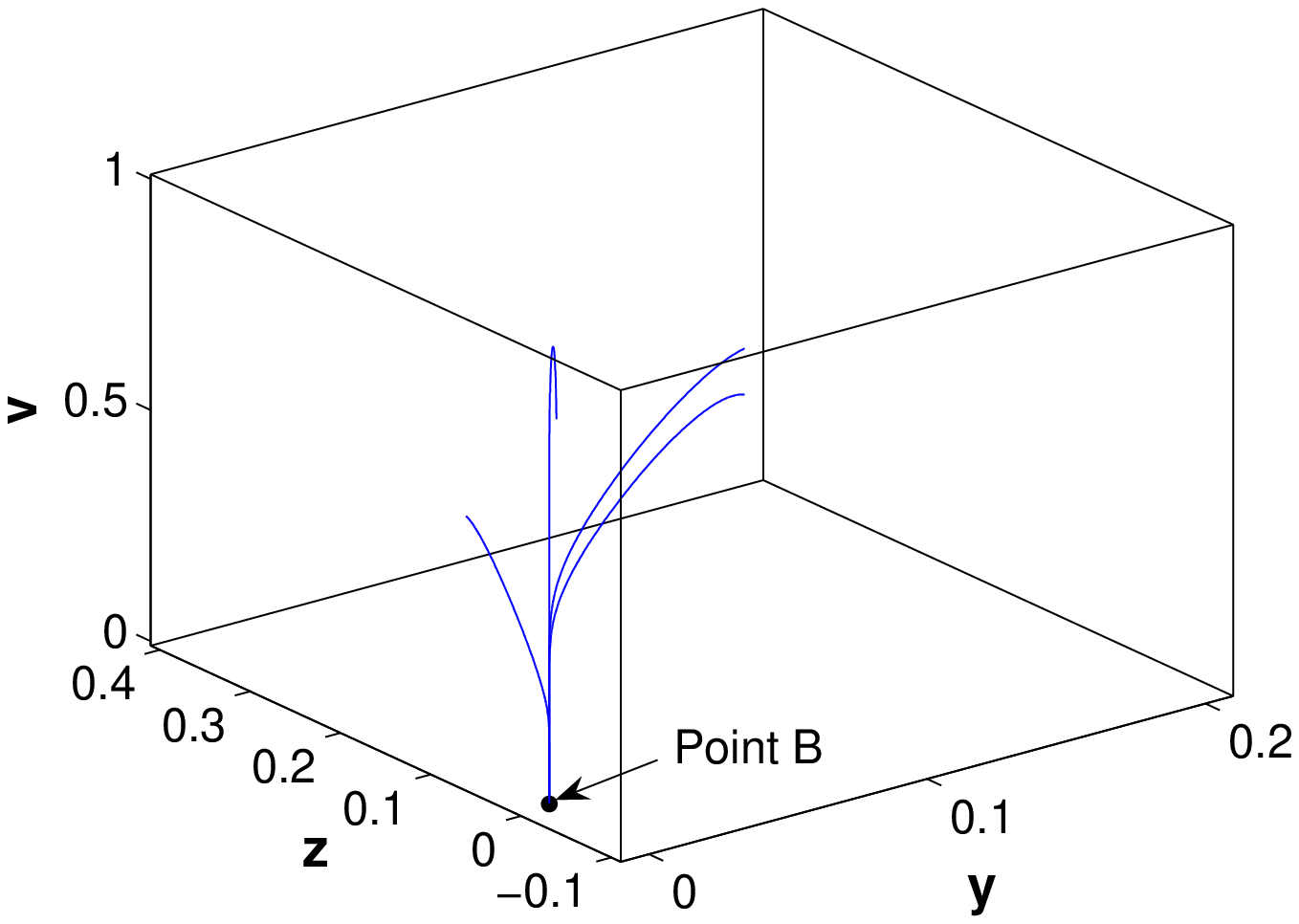}
\end{center}
\caption{Three-dimensional phase space for the interacting model II
with $w_{d}=-1.2$ in classical Einstein cosmology. The left is for
the phase space of ($x,y,z$) and the right is for the phase space of
($y,z,v$).} \label{Fig.1}
\end{figure}

\begin{figure}[tbp]
\begin{center}
\includegraphics[width=0.3\textwidth]{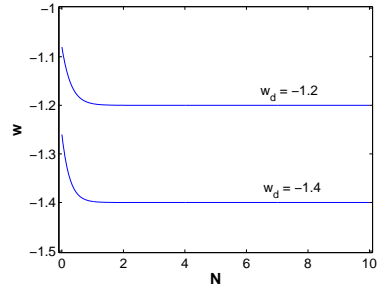}
\end{center}
\caption{The evolution of the EoS of total cosmic fluid $w$ for the
interacting model II with $\beta=10^{-6}$ in LQC (the same as in
classical Einstein cosmology). } \label{Fig.1}
\end{figure}

\begin{figure}[tbp]
\begin{center}
\includegraphics[width=0.3\textwidth]{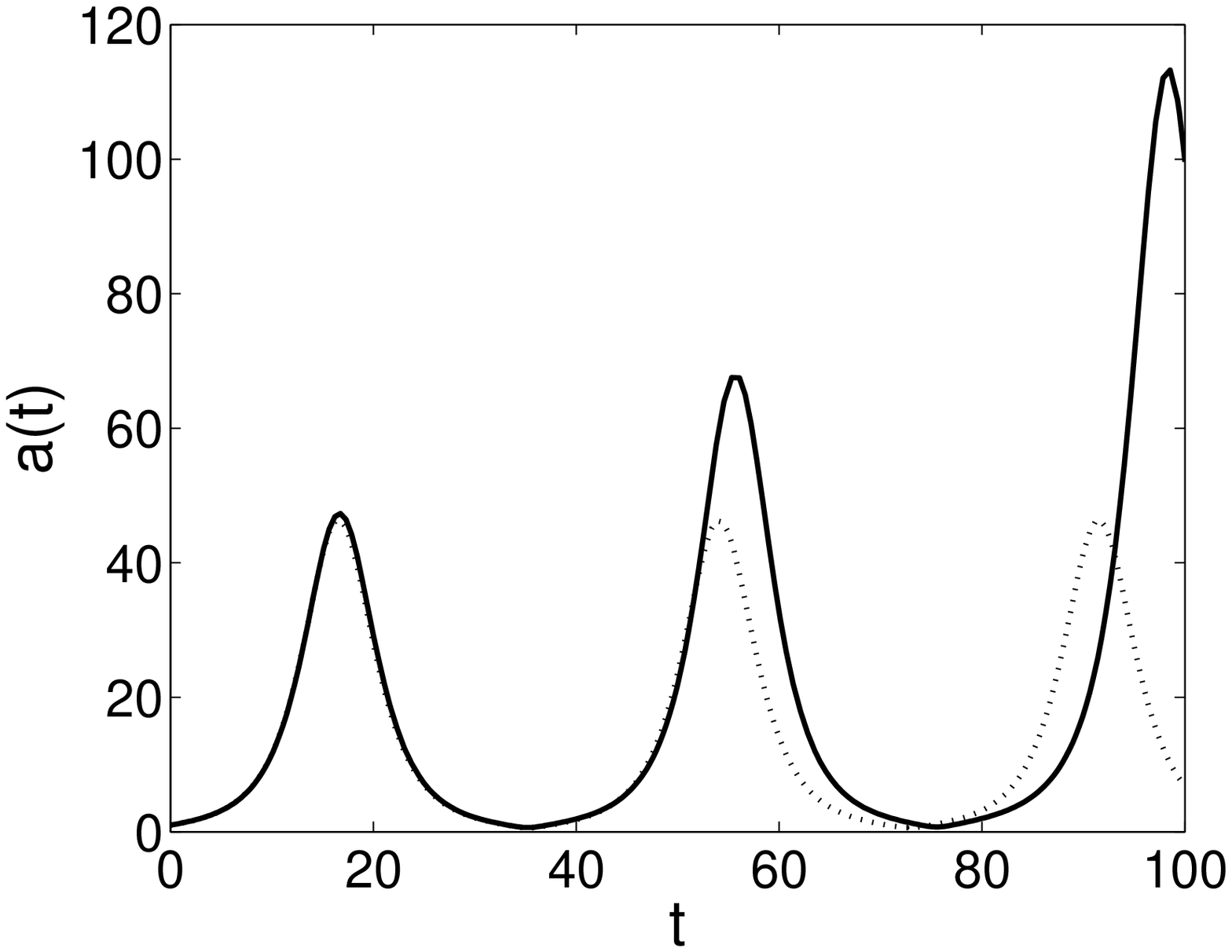}%
\includegraphics[width=0.3\textwidth]{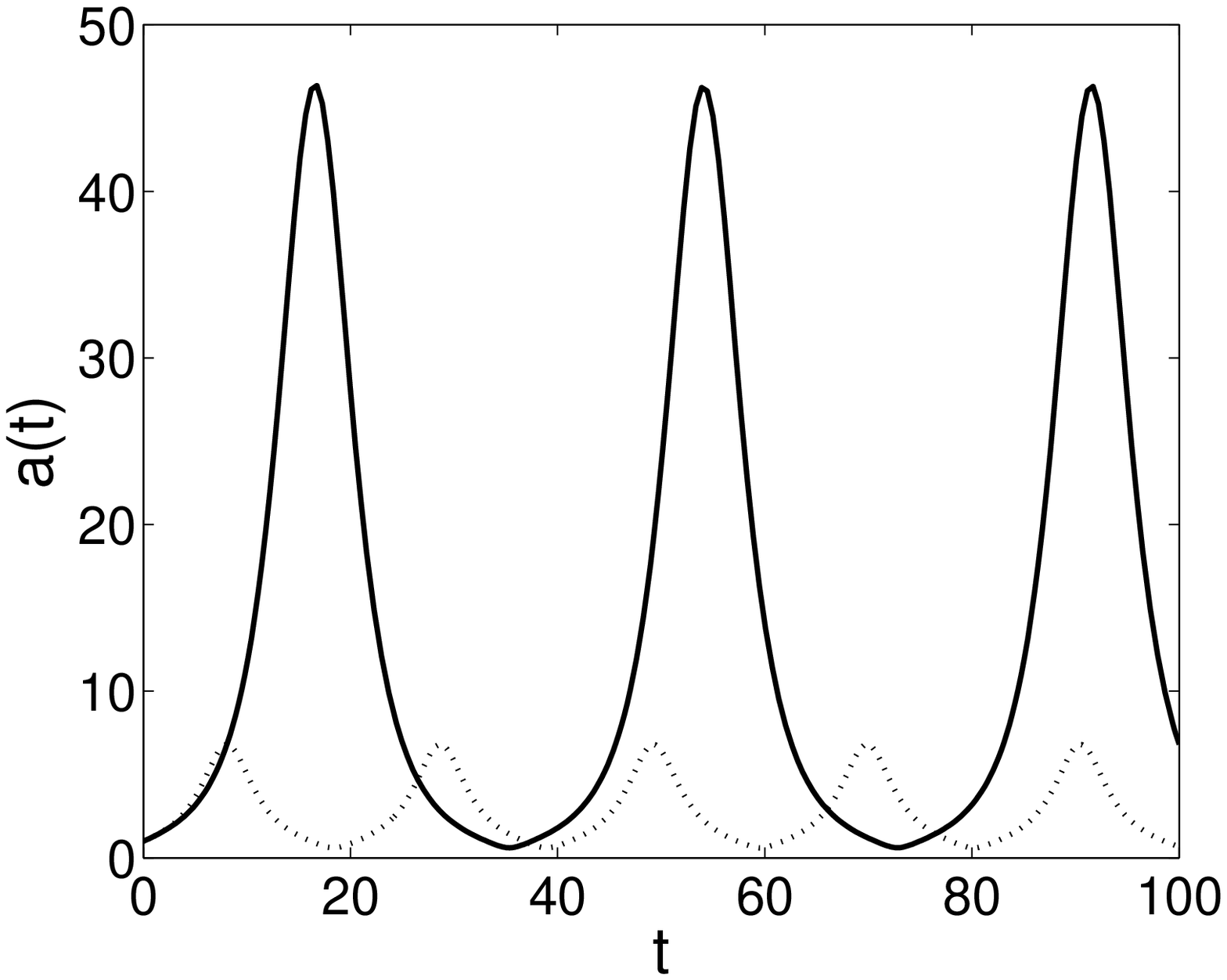}
\end{center}
\caption{The plots of scalar factor $a$ as a function of time for
the interacting model II. The left plot is for $w_{d}=-1.2$, in
which the solid and dotted lines denote the cases of $\beta=10^{-2}$
and $\beta=10^{-6}$, respectively. The right is for $\beta=10^{-6}$,
in which the solid and dotted lines represent $w_{d}=-1.2$ and
$w_{d}=-1.4$, respectively.} \label{Fig.1}
\end{figure}

\begin{figure}[tbp]
\begin{center}
\includegraphics[width=0.3\textwidth]{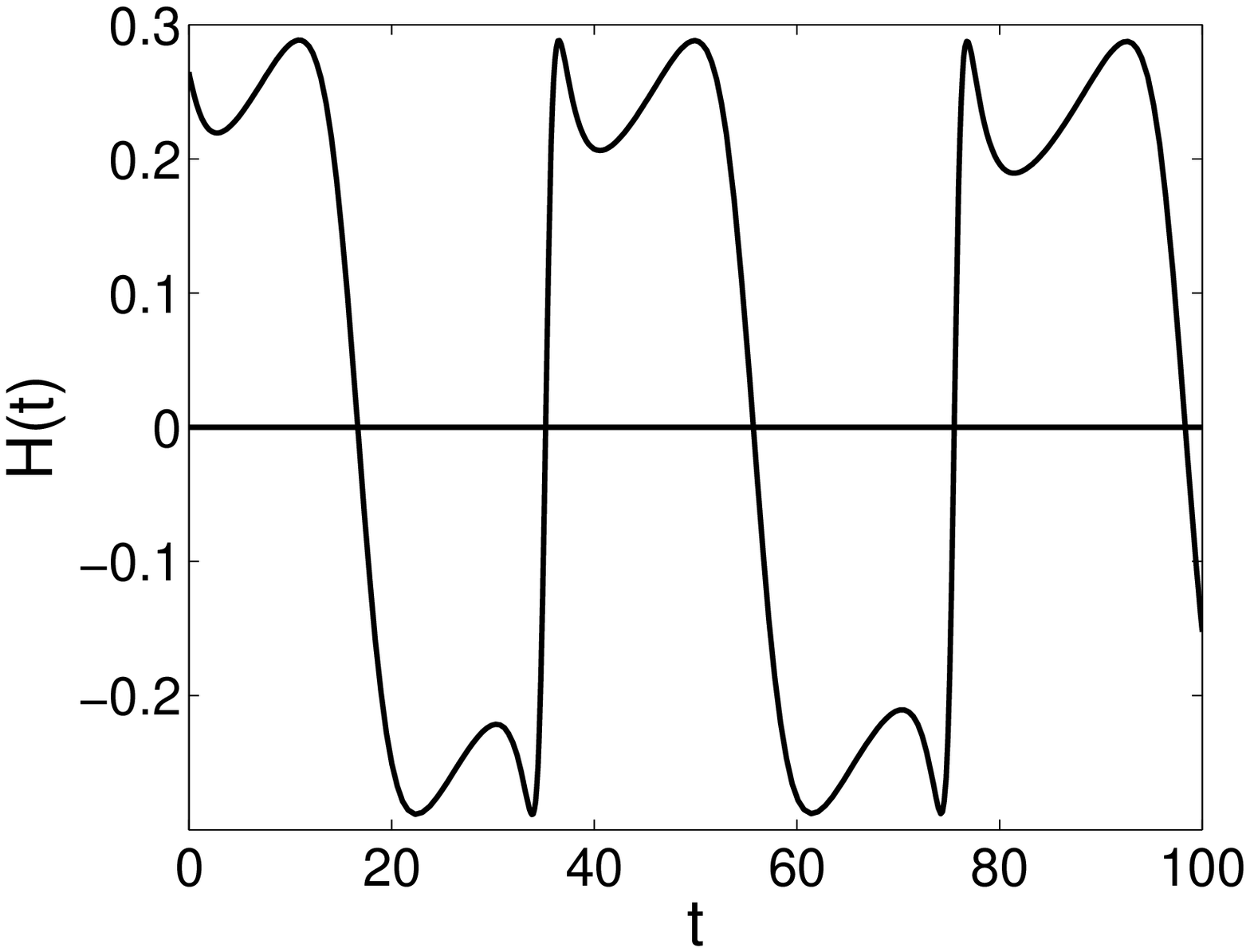}%
\includegraphics[width=0.3\textwidth]{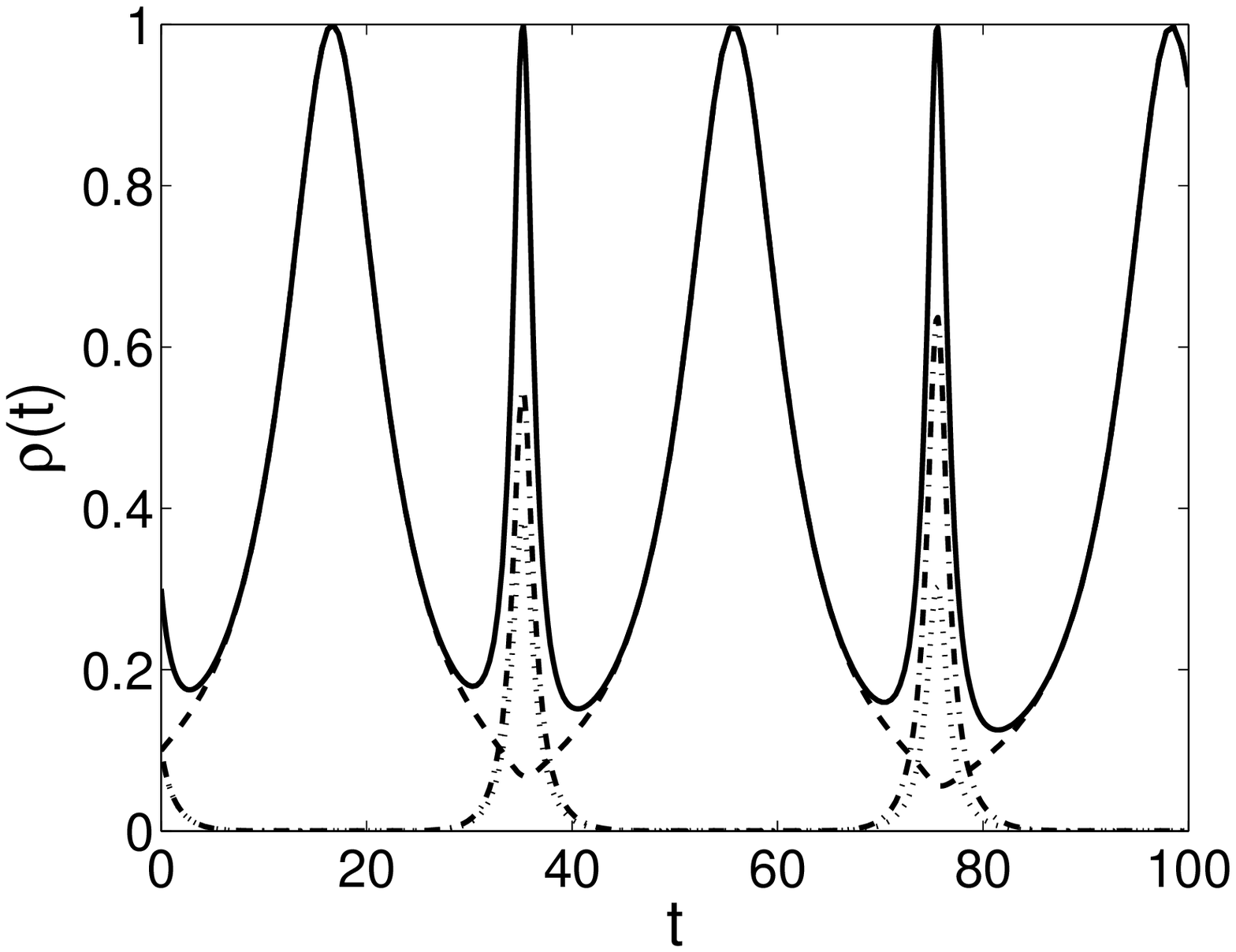}
\end{center}
\caption{The evolution of the Hubble parameter $H$ and the energy
density $\rho$ with respect to time for $w_{d}=-1.2,\beta=10^{-2}$.
The solid, dashed, dash-dotted and dotted lines correspond to
$\rho$, $\rho_{d}$, $\rho_{m}$ and $\rho_{b}$, respectively. }
\label{Fig.1}
\end{figure}

\begin{figure}[tbp]
\begin{center}
\includegraphics[width=0.3\textwidth]{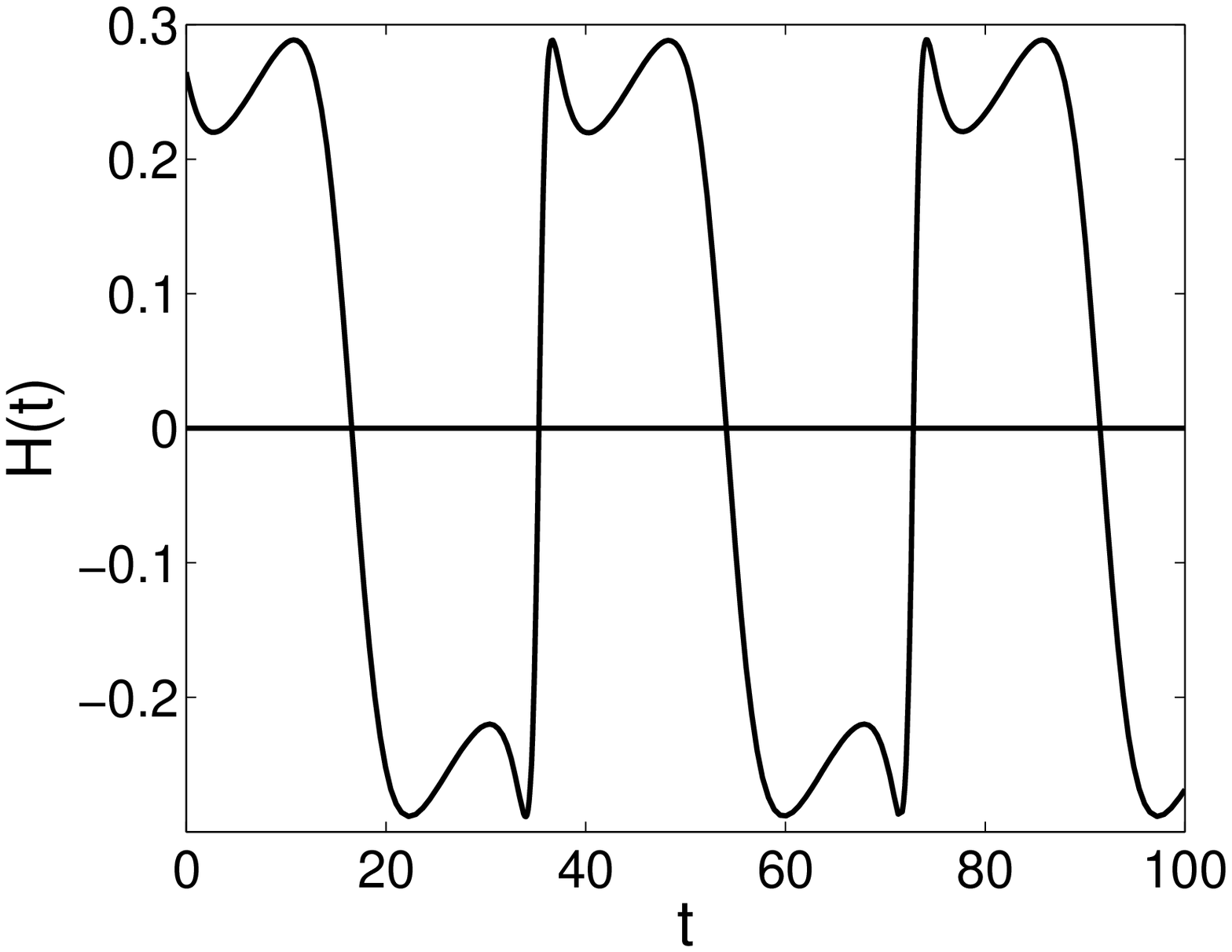}%
\includegraphics[width=0.3\textwidth]{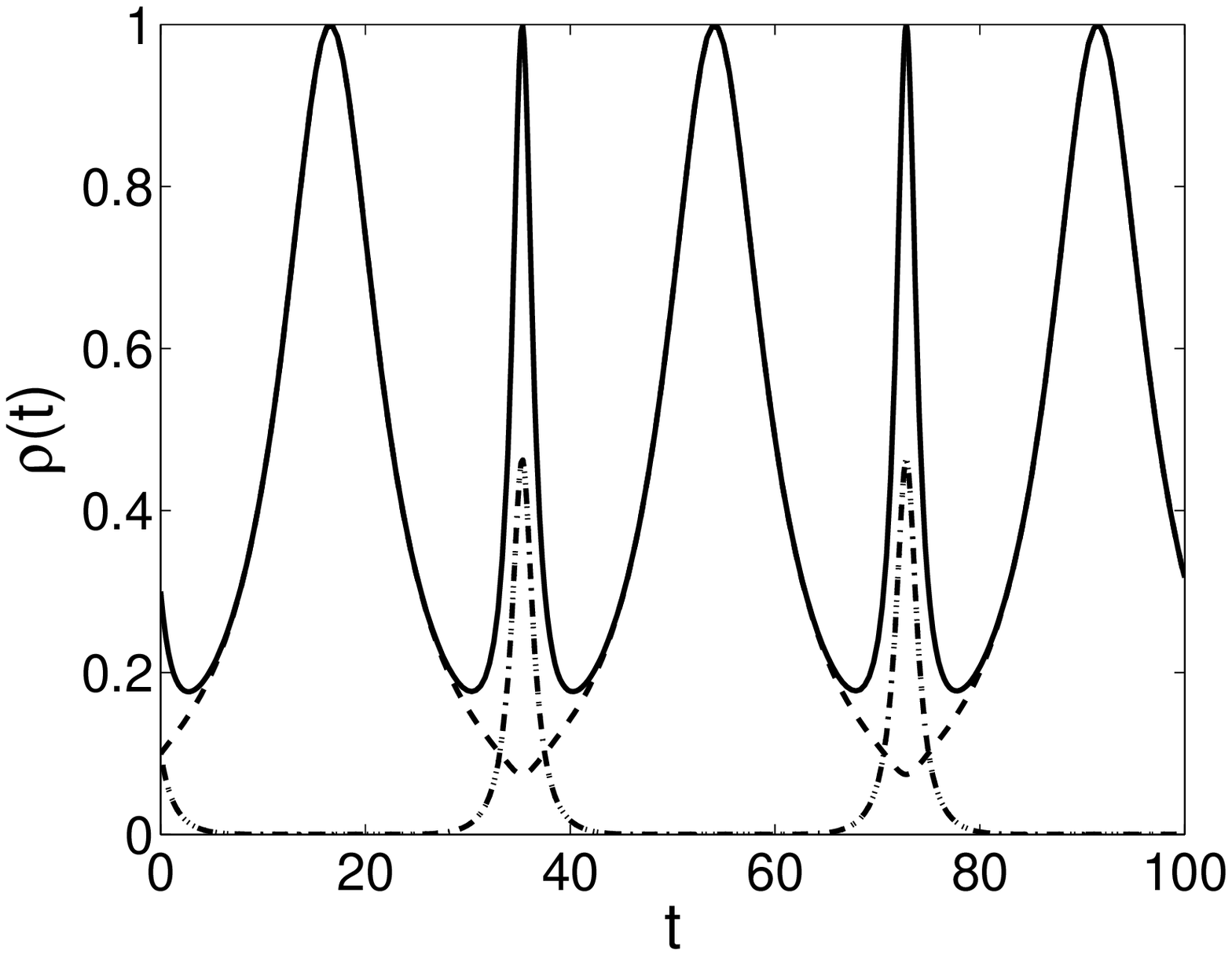}
\end{center}
\caption{The evolution of the Hubble parameter $H$ and the energy
density $\rho$ with respect to time for $w_{d}=-1.2,\beta=10^{-6}$.
The solid, dashed, dash-dotted and dotted lines correspond to
$\rho$, $\rho_{d}$, $\rho_{m}$ and $\rho_{b}$, respectively. }
\label{Fig.1}
\end{figure}

\begin{figure}[tbp]
\begin{center}
\includegraphics[width=0.3\textwidth]{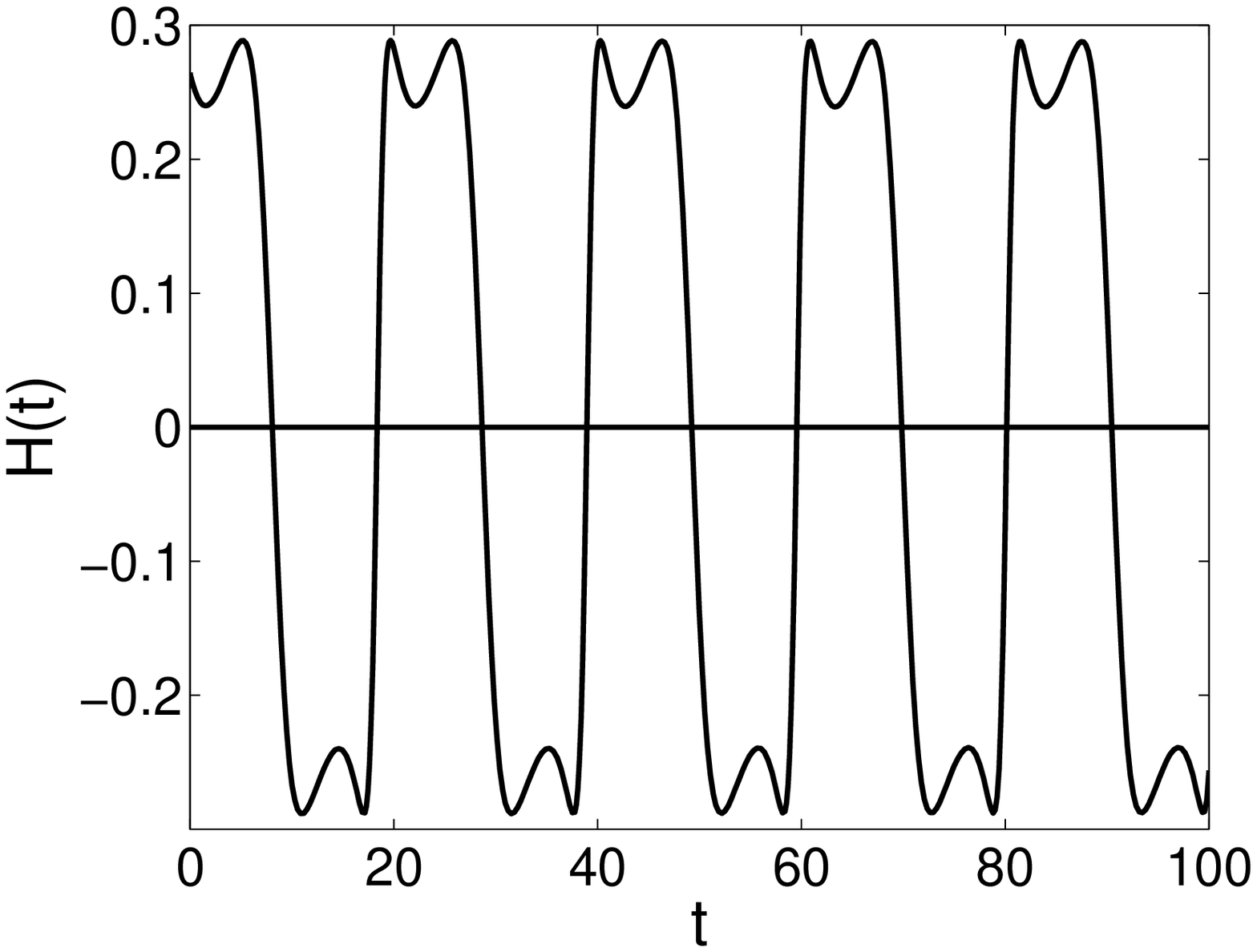}%
\includegraphics[width=0.3\textwidth]{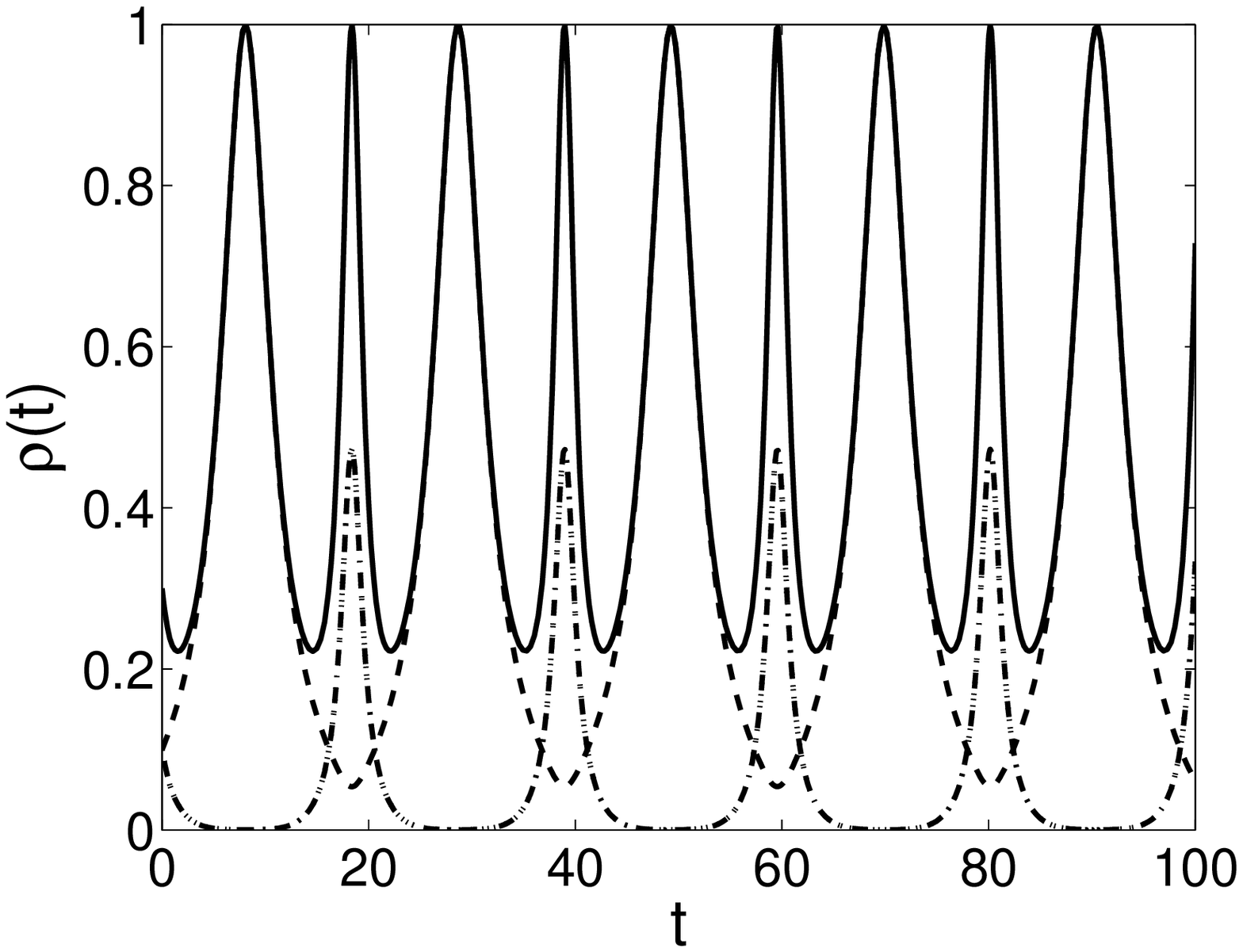}
\end{center}
\caption{The evolution of the Hubble parameter $H$ and the energy
density $\rho$ with respect to time for $w_{d}=-1.4,\beta=10^{-6}$.
The solid, dashed, dash-dotted and dotted lines correspond to
$\rho$, $\rho_{d}$, $\rho_{m}$ and $\rho_{b}$, respectively. }
\label{Fig.1}
\end{figure}

\end{document}